\documentclass[11pt,a4paper]{article}
\pdfoutput=1

\usepackage{jheppub}
\usepackage{feynmf}
\usepackage{pifont}
\usepackage{amssymb}
\usepackage{graphicx}
\usepackage{verbatim}

\def \be {\mathbf{E}}

\def \br {\mathbf{r}}

\def \bq {\mathbf{q}}

\def \cf {C_F}
\def \nc {N_c}

\def \bk {\mathbf{k}}
\def \mbk {\vert\bk\vert}
\def \bp {\mathbf{p}}

\newcommand{\Tint}[1]{{\hbox{$\sum$}\!\!\!\!\!\!\!\int\,}_{\!\!\!\!\raise-0.9ex\hbox{$\scriptstyle{#1}$}}}

\def \mbq {\vert\bq\vert}

\def\siml{{\ \lower-1.2pt\vbox{\hbox{\rlap{$<$}\lower6pt\vbox{\hbox{$\sim$}}}}\ }}
\def\simg{{\ \lower-1.2pt\vbox{\hbox{\rlap{$>$}\lower6pt\vbox{\hbox{$\sim$}}}}\ }}

\def \be {\mathbf{E}}

\def \br {\mathbf{r}}

\def \bq {\mathbf{q}}

\def \bk {\mathbf{k}}
\def \mbk {\vert\bk\vert}
\def \bp {\mathbf{p}}

\def \mbq {\vert\bq\vert}

\def \als {\alpha_{\mathrm{s}}}

\def \m2   {\mu^{2 \epsilon}}

\def\siml{{\ \lower-1.2pt\vbox{\hbox{\rlap{$<$}\lower6pt\vbox{\hbox{$\sim$}}}}\ }}
\def\simg{{\ \lower-1.2pt\vbox{\hbox{\rlap{$>$}\lower6pt\vbox{\hbox{$\sim$}}}}\ }}
\def\lqcd{\Lambda_\mathrm{QCD}}

\def\nn {\nonumber}

\title{Thermal width and quarkonium dissociation by inelastic parton scattering}
\author[a]{Nora Brambilla}
\author[a]{Miguel \'Angel Escobedo}
\author[b]{Jacopo Ghiglieri}
\author[a]{Antonio Vairo}
\affiliation[a]{Physik-Department, Technische Universit\"at M\"unchen,\\
James-Franck-Str. 1, 85748 Garching, Germany}
\affiliation[b]{Department of Physics, McGill University, 3600 rue University, Montr\'eal QC H3A 2T8, Canada}

\emailAdd{nora.brambilla@ph.tum.de}
\emailAdd{miguel.escobedo@ph.tum.de}
\emailAdd{jacopo.ghiglieri@physics.mcgill.ca}
\emailAdd{antonio.vairo@ph.tum.de}

\preprint{TUM-EFT 27/11}

\abstract{In a weak-coupling effective field theory framework we study
quarkonium dissociation induced by inelastic scattering with partons in the
medium. This is the dominant dissociation process for temperatures such that
the Debye mass is larger than the binding energy. We evaluate the
dissociation cross section and the corresponding thermal decay width. At
leading order we derive a convolution formula relating the two, which is
consistent with the optical theorem and QCD at finite temperature. Bound
state effects are systematically included. They add contributions to the
cross section and width that are beyond a quasi-free approximation, whose
validity is critically reviewed. For temperatures such that the Debye mass
is smaller than the binding energy, the dominant dissociation mechanism is
gluo-dissociation consisting in quarkonium dissociation induced by the
absorbtion of a gluon from the medium. We calculate the gluo-dissociation
cross section and width at next-to-leading-order accuracy.}


\begin{document}

\maketitle

\section{Introduction} 
Heavy-quarkonium suppression, first proposed in~\cite{Matsui:1986dk} as a signal of deconfinement, 
has been observed at SPS, RHIC and recently at LHC~\cite{Brambilla:2004wf,Brambilla:2010cs,Chatrchyan:2012lxa,Abelev:2012rv}. 
Although the current understanding is that the observed quarkonium suppression 
cannot be explained by cold nuclear matter effects alone~\cite{Brambilla:2010cs},   
the hot nuclear matter mechanism responsible for it is still under investigation. 
In~\cite{Matsui:1986dk}, it was suggested that heavy quark-antiquark ($Q\overline{Q}$) bound states 
dissociate in a hot thermal bath because of colour screening of the $Q\overline{Q}$ potential induced by the medium. 
In~\cite{Laine:2006ns}, another dissociation mechanism was identified in the Landau-damping phenomenon. 
Implications of the Landau-damping mechanism on the quarkonium dynamics, and anisotropic generalizations thereof, 
can be found in~\cite{Burnier:2007qm,Petreczky:2010tk,Strickland:2011mw,Strickland:2011aa}.

In the context of real-time thermal field theory, quarkonium in a thermal bath may 
be studied by taking advantage of the non-relativistic and thermal energy scales that characterize the system.
The scales typical of a non-relativistic bound state are the heavy-quark mass $m$, 
the momentum transfer or inverse radius $1/r \sim mv$ and the binding energy $E\sim mv^2$.
Because $v \ll 1$ is the relative velocity of the heavy quarks, these scales are hierarchically
ordered:  $m\gg mv\gg mv^2$. The thermal bath is charactetized by a temperature $T$ and a Debye screening mass $m_D$. 
In a weakly-coupled plasma, $m_D\sim gT$ and also the thermal scales are ordered: $T \gg m_D$.
Integrating out systematically degrees of freedom associated with the highest energy scale leads 
to a hierarchy of low-energy effective field theories (EFTs)~\cite{Escobedo:2008sy,Brambilla:2008cx}. 
The ultimate of these EFTs can be interpreted as a finite-temperature version 
of potential non-relativistic QCD (pNRQCD)~\cite{Pineda:1997bj,Brambilla:1999xf}. 
Potential non-relativistic QCD describes the quarkonium dynamics through potentials and low-energy interactions.
Thermal corrections affect both the real and the imaginary parts 
of the potentials. Thermal corrections to the real part of the colour-singlet 
potential may lead to the colour-screening phenomenon described by Matsui and Satz, 
whereas thermal corrections to the imaginary part of the colour-singlet potential induce a thermal width. 
Several mechanisms may contribute to the imaginary part of the potential. 
One of these is precisely the Landau-damping mechanism identified in~\cite{Laine:2006ns}. 
In~\cite{Escobedo:2008sy,Laine:2008cf}, it was shown that the thermal width induced 
by the Landau-damping phenomenon is the principal source of quarkonium 
dissociation at weak coupling, responsible for keeping the quarkonium dissociated even at temperatures where 
colour screening on the real part of the potential has faded away. 
Moreover, lattice studies have shown that the potential 
may have a sizeable imaginary part also at strong coupling~\cite{Laine:2007qy,Rothkopf:2011db,Burnier:2012az}.

In a non-EFT framework, quarkonium decay widths induced by scattering with the medium constituents 
have been studied since long time (see for instance~\cite{Mocsy:2007jz,Rapp:2008tf,Kluberg:2009wc,Rapp:2009my} 
and references therein). At leading order, two different dissociation mechanisms were identified:
\emph{gluo-dissociation}~\cite{Kharzeev:1994pz,Xu:1995eb} and
\emph{dissociation by inelastic parton scattering}~\cite{Grandchamp:2001pf,Grandchamp:2002wp}. 
In the former case, the bound state absorbs a sufficiently energetic 
gluon of the medium and dissociates into an unbound colour-octet $Q\overline{Q}$ pair.
The gluon is physical, in the sense that its momentum is either light-like 
or time-like if it acquires an effective mass propagating through the medium.
In the latter  case a light parton of the medium, gluon or quark, scatters off the bound state
by exchanging gluons, resulting again in its dissociation into an unbound colour octet. 
The momentum of the exchanged gluon is in this case space-like.
In both cases, the decay widths were obtained by convoluting the $T=0$ cross section
for the scattering process, possibly with some ad-hoc finite-temperature
modifications, with the thermal distribution of the incoming parton.

In the EFT framework, thermal decay widths have been investigated over 
a wide range of temperatures~\cite{Escobedo:2008sy,Brambilla:2008cx,Brambilla:2010vq}. 
In~\cite{Brambilla:2008cx}, two mechanisms contributing at leading order 
to the quarkonium decay width 
were identified: \emph{singlet-to-octet thermal breakup} and \emph{Landau damping}. 
In the power-counting of the EFT, it was shown that the latter dominates over the former 
as long as $m_D\gg E$. These two mechanisms are not independent from those identified 
earlier without EFT methods. In terms of elementary processes, 
the singlet-to-octet thermal breakup corresponds to gluon-dissociation
and the Landau-damping mechanism to dissociation by inelastic parton scattering~\cite{Riek:2010py}.
Beyond leading order the two mechanisms are intertwined and distinguishing 
between them becomes unphysical, whereas the physical quantity is the total width. 

The equivalence of the singlet-to-octet thermal breakup process to gluo-dissociation 
has been analyzed at leading order in~\cite{Brambilla:2011sg}. There the singlet-to-octet thermal breakup width 
and the gluo-dissociation cross section have been computed for several temperature re\-gi\-mes.
The decay width in the regime $mv \gg T \gg E \gg m_D$,  which has been suggested to be of relevance for 
$\Upsilon(1S)$ suppression at the LHC~\cite{Vairo:2010bm},
agrees with the one previously calculated in~\cite{Brambilla:2010vq}. 
Moreover it is consistent with the lattice QCD findings of~\cite{Aarts:2011sm}.
The gluo-dissociation cross section agrees in the large-$N_c$ limit 
with the cross section derived in~\cite{Bhanot:1979vb}, which is often used in the literature.
The gluo-dissociation cross section has been confirmed by~\cite{Brezinski:2011ju}. 

In this paper, we will perform a similar analysis for what concerns the relation 
between the Landau-damping mechanism and the dissociation by inelastic parton scattering.
We will compute in an EFT framework the dissociation cross section 
and the corresponding thermal width in different temperature regimes and relate 
the results with the existing literature.
The cross section for dissociation by inelastic parton scattering was computed in~\cite{Combridge:1978kx} 
neglecting bound-state effects, and more recently (but in a different validity region) in~\cite{Song:2005yd}.
The relevance of this process for heavy-ion collisions was realized in~\cite{Grandchamp:2001pf} 
and since then phenomenological expressions for the quarkonium decay width due 
to dissociation by inelastic parton scattering in the medium have been widely used in the literature 
(see~\cite{Grandchamp:2002wp,Grandchamp:2005yw,Park:2007zza,Zhao:2010nk,Emerick:2011xu,Song:2011ev}). 
As this process gives the dominant contribution to the decay width for
$m_D\gg E$, which is the temperature regime at which quarkonium dissociates, 
validating those expressions from QCD is of utmost phenomenological relevance. 
In fact, we will show that the convolution formula commonly used in the literature to
connect the dissociation cross section to the decay width does not follow 
from the optical theorem applied to QCD at finite temperature 
and we will suggest a different one. Furthermore, we will argue that neglecting 
bound-state effects is not a valid approximation in a weak-coupling setting below the 
quarkonium dissociation temperature.

The paper is organized as follows. In section~\ref{sec:gen}, we give
some general arguments based on the optical theorem in thermal field theory
and on effective field theories to derive a formula relating the 
dissociation cross section to the decay width. In section~\ref{sec_EFT}, 
we illustrate the basics of pNRQCD. In section~\ref{sec_tggr}, we study
the cross section and width in the temperature regime $T \gg mv \sim m_D$ 
and in section~\ref{sec_tsimr} we consider the regime $T \sim m v \gg m_D$. 
In section~\ref{sec_rggt}, we focus on the temperature regime $m v\gg T\gg m_D\gg E$, 
where a multipole expansion in the temperature is possible. 
We obtain a colour dipole cross section and we study the
validity region of this approximation. In section~\ref{sec_eggmd}, 
we analyze the temperature regime $m v \gg T \gg E \gg m_D$, where,  
as we shall show, both gluo-dissociation and dissociation by parton scattering get a contribution 
from the same EFT diagram. Hence, we compute next-to-leading-order (NLO) corrections 
to the gluo-dissociation cross section. Finally, in section~\ref{sec_concl} we draw some conclusions.

\section{General considerations on dissociation by inelastic parton scattering}
\label{sec:gen} 
\begin{figure}[ht]
	\begin{center}
\includegraphics[scale=0.55]{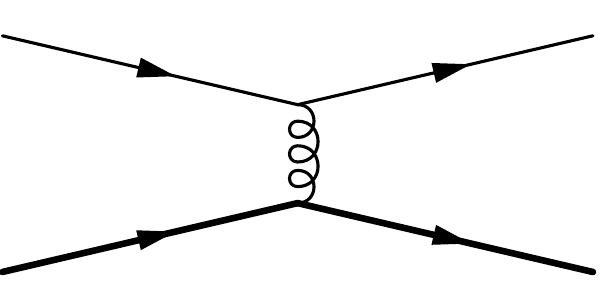}\,\,\,\,\includegraphics[scale=0.55]{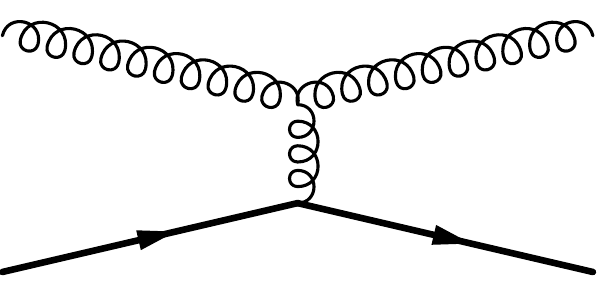}\,\,\,\,
\includegraphics[scale=0.55]{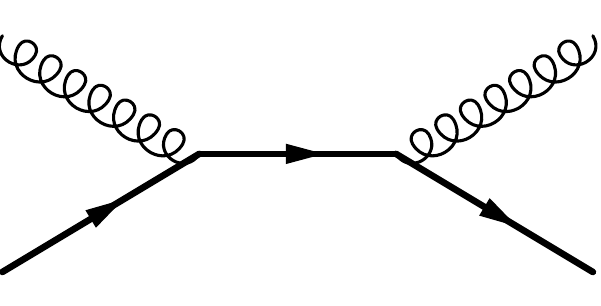}\,\,\,\,\includegraphics[scale=0.55]{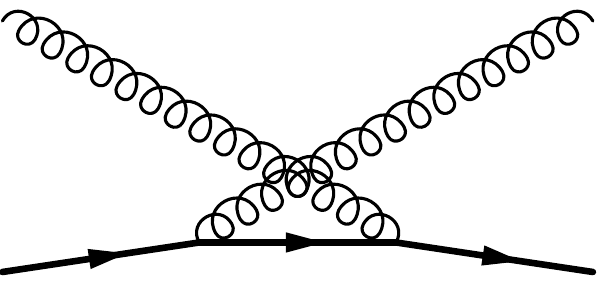}
\end{center}
\caption{Tree level diagrams contributing to  $p\,Q\to p\,Q$, where $p$ is a parton and $Q$ a heavy quark.  
 Thick lines stand for heavy quarks, curly lines for gluons and thin lines for light quarks.}
\label{fig:com}
\end{figure}

It has been suggested in~\cite{Grandchamp:2001pf} and used in most of the following 
literature on the subject that the width $\Gamma_{\rm HQ}$ for dissociation by inelastic parton
 scattering of a heavy quarkonium, ${\rm HQ}$, at rest with respect to the thermal bath 
could be expressed by the convolution formula
\begin{equation}
\Gamma_{\rm HQ}=\sum_p\int_{q_\mathrm{min}}\frac{d^3q}{(2\pi)^3}\,f_p(q)\,\sigma_{p}^{\rm HQ}(q)\,,
\label{eq:facw}
\end{equation} 
where the sum runs over the different species of incoming light partons, $p$,  
with momentum $q=\mbq$, and the distribution functions, $f_p$, 
are the Bose--Einstein distribution $n_\mathrm{B}(q)=1/(\exp(q^0/T)-1)$ 
for gluons and the Fermi--Dirac distribution $n_\mathrm{F}(q)=1/(\exp(q^0/T)+1)$ 
for light quarks. The momentum $q_\mathrm{min}$ is the minimum incoming momentum 
necessary to dissociate the bound state. 
The distribution functions are convoluted with $\sigma_{p}^{\rm HQ}$, a quantity identified with 
the parton-heavy-quarkonium dissociation cross section in the medium.
In~\cite{Grandchamp:2001pf} and related literature, $\sigma_{p}^{\rm HQ}$ 
has been approximated by $2\sigma_{p}^{Q}$, where $\sigma_{p}^{Q}$ is the cross section 
of the zero-temperature process $p\,Q\to p\,Q$.
Because this approximation neglects bound-state effects, it is called quasi-free.\footnote{
In the literature, dissociation by inelastic parton scattering is sometimes
called for conciseness quasi-free dissociation, after the approximation used for its computation.
} 
The scattering process $p\,Q\to p\,Q$ receives at leading order in perturbation theory 
contributions from the four diagrams shown in figure~\ref{fig:com};
these were computed in~\cite{Combridge:1978kx}. 
It is the cross section computed in~\cite{Combridge:1978kx} that is commonly used in eq.~\eqref{eq:facw}.
Besides the distribution functions, additional thermal effects are usually added to \eqref{eq:facw} in the form 
of momentum-independent thermal masses affecting the dispersion relations 
and propagators of the light partons. Thermal masses also provide a cut off 
for the infrared divergences typically affecting the forward scattering amplitude. 

Our purpose is to scrutinize eq.~\eqref{eq:facw} and the quasi-free approximation 
in the light of QCD at finite temperature. We start by making some general considerations 
about the form of the dissociation formula. A quantitative treatment 
in a weak-coupling regime will be developed in the next sections.
The dissociation process we are considering happens when partons in the thermal 
bath, i.e. partons with momentum and energy of the order of the temperature, 
scatter off the quarkonium and dissociate it by exchanging space-like gluons.
First, we observe that by performing the calculation in Coulomb gauge, we only need to take into account 
transverse gluons, $A_i$, as external gauge fields. The reason is that in Coulomb gauge temporal gluons, $A_0$, 
do not thermalize at the scale~$T$; moreover their spectral density 
vanishes and hence they do not contribute to cut diagrams. 
The Coulomb gauge is therefore a convenient gauge for the calculation, and we will adopt it for the rest of the paper. 
Next, we assume that the heavy-quark mass is much larger that the temperature: $m \gg T$. 
This implies that one can integrate out hard modes with energy and momentum of order $m$ from QCD neglecting thermal effects, 
and replace QCD by non-relativistic QCD (NRQCD)~\cite{Caswell:1985ui,Bodwin:1994jh} as the fundamental theory. 
In NRQCD, the leading interaction between heavy quarks and gluon fields $A_i$ is encoded 
in dimension-five operators; each of these interactions brings a suppression factor proportional to $T/m$.
We conclude that in Coulomb gauge the last two diagrams in figure~\ref{fig:com} are at least suppressed 
by a factor $T/m$ with respect to the first two and can be neglected at leading order.

Whenever the momentum carried by the gluon coupled to the heavy quark is larger than 
or of the same order as $mv$, based on the same argument given in the previous paragraph 
we can argue that the leading-order contribution of the first two diagrams in figure~\ref{fig:com}
comes from the coupling of a temporal gluon to the heavy quark.
This case will be relevant for section~\ref{sec_tggr} and~\ref{sec_tsimr}. 
Whenever the momentum carried by the gluon coupled to the heavy quark is smaller than $mv$, but larger than 
the binding energy, the contribution of the temporal gluons cancels at leading-order in the multipole expansion 
in the square of the amplitude of the first two diagrams of figure~\ref{fig:com} and the corresponding antiquark diagrams.
This was noted in~\cite{Brambilla:2008cx} and will be shown again in the following.
Nevertheless, also in this case we may neglect at leading order the coupling of a transverse gluon to the heavy quark. 
The reason is that its contribution is proportional to the binding energy of the quarkonium, and this is subleading 
with respect to the momentum carried by the gluon. 
This case will be relevant for most of the temperature regions studied in the paper.
Finally, whenever the momentum carried by the gluon coupled to the heavy quark is of the same order as the binding energy, 
the contribution coming from a transverse gluon coupled to the heavy quark 
will be at next-to-leading order in the multipole expansion as important as the one coming from a temporal gluon 
and we will need to consider them both. This case will be relevant only for section~\ref{sec_eggmd}.

Since in all cases only the first two diagrams in figure~\ref{fig:com} contribute at non-vanishing leading order, 
it follows from the optical theorem that the dissociation cross section is proportional to the imaginary 
part of $iD^{\mu\nu}$, where $D^{\mu\nu}$ is the gluon propagator. 
The gluon propagator may be computed in the so-called real-time formalism of thermal field theory (see e.g.~\cite{Lebellac}). 
A feature of this formalism is that the degrees of freedom double: while external particles are only of 
type ``1'', i.e. they live on the time-ordered branch of the Schwinger--Keldysh contour, 
in loops one has to consider also particles of type ``2'', i.e. particles located 
on the anti-time-ordered branch. It has been shown in~\cite{Brambilla:2008cx,Ghiglieri:2012rp} that 
under the condition $m \gg T$ heavy quarks do not thermalize up to exponentially suppressed contributions 
and can be treated as external probes. Hence, all vertices involving 
heavy quarks are of type ``1'' and, as a consequence, one needs to consider only the imaginary part 
of the ``11'' component of the gluon propagator coupled to the heavy quark.
In general, this can be written in terms of an advanced ($A$), retarded ($R$) and symmetric ($S$) propagator as  
\begin{equation}
D^{\mu\nu}_{11}(k_0,k)=\frac{1}{2}\left[D^{\mu\nu}_R(k_0,k)+D^{\mu\nu}_A(k_0,k)+D^{\mu\nu}_S(k_0,k)\right]\,,
\end{equation}
where throughout the paper italic letters refer to the modulus of the spatial momentum, i.e. $k=\mbk$.
Since $(iD^{\mu\nu}_R)^*=iD^{\mu\nu}_A$, the discontinuity $iD^{\mu\nu}_{11} - (iD^{\mu\nu}_{11})^*$ is equal to $iD^{\mu\nu}_S$, 
which is in turn related to the retarded and advanced propagators through  
$D^{\mu\nu}_S(k_0,k)=[1 + 2 n_{\rm B}(\vert k_0\vert)]\,\mathrm{sgn}(k_0)\,[D^{\mu\nu}_R(k_0,k)-D^{\mu\nu}_A(k_0,k)]$.
The resummed retarded/advanced propagator is obtained by resumming the retarded/advanced gluon self energy,
$\Pi^{\mu\nu}_{R,A}$. The gluon polarization tensor satisfies relations similar to those valid for the gluon propagator.
In particular, the ``11'' component of the self energy can be written as 
\begin{equation}
\Pi^{\mu\nu}_{11}(k_0,k)=\frac{1}{2}\left[\Pi^{\mu\nu}_R(k_0,k)+\Pi^{\mu\nu}_A(k_0,k)+\Pi^{\mu\nu}_S(k_0,k)\right]\,.
\end{equation}
From $(\Pi^{\mu\nu}_{R})^* = \Pi^{\mu\nu}_{A}$, it follows that the discontinuity 
$\Pi^{\mu\nu}_{11} - (\Pi^{\mu\nu}_{11})^*$ is equal to $\Pi^{\mu\nu}_S$, since we also have that 
$\Pi^{\mu\nu}_S(k_0,k)=[1 + 2n_{\rm B}(\vert k_0\vert)]\,\mathrm{sgn}(k_0)\,[\Pi^{\mu\nu}_R(k_0,k)$ $-\Pi^{\mu\nu}_A(k_0,k)]$.
In summary, the imaginary part of $iD^{\mu\nu}_{11}$ for spacelike momenta is proportional to $iD^{\mu\nu}_S$, which, in turn, 
is proportional to~$\Pi^{\mu\nu}_S$. In relation to the dissociation by inelastic parton scattering
 indeed only $\Pi^{\mu\nu}_S$ matters, 
because contributions coming from the poles of the propagator are not space-like.  

For the purpose of scrutinizing eq.~\eqref{eq:facw}, we will 
concentrate now on the case of a temporal gluon propagator in Coulomb gauge with 
incoming momentum $k \gg k_0$. This is the only case needed in all sections of the paper
  with the exception of section~\ref{sec_eggmd}.
The resummed temporal gluon propagator in Coulomb gauge is given by 
\begin{equation}
D^{00}_{R,A}(k_0,k)=\frac{i}{k^2+\Pi^{00}_{R,A}(k_0,k)}\,.
\end{equation}
It follows that the temporal symmetric gluon propagator can be written as 
\begin{equation}
D^{00}_S(k_0,k)= -i\frac{\Pi^{00}_S(k_0,k)}{(k^2+\Pi^{00}_R(k_0,k))(k^2+\Pi^{00}_A(k_0,k))}\,,
\label{deltaspi}
\end{equation}
which makes clear that the imaginary part of $iD^{00}_{11}$ for space-like momenta comes from $\Pi^{00}_S$.
At one loop, $\Pi^{00}_S(k_0,k)$ is the sum of a light-quark, $\Pi^{00}_{S,\,q}(k_0,k)$, and a gluon contribution,  $\Pi^{00}_{S,\,g}(k_0,k)$. 
We will consider them for momenta $k \gg k_0$.
The symmetric gluon self energy can be computed by means of the cutting rules at finite temperature 
introduced in~\cite{Kobes:1985kc,Kobes:1986za} (see also~\cite{Bedaque:1996af,Gelis:1997zv}).
According to them, the quark contribution to the longitudinal polarization tensor reads
\begin{equation}
\Pi^{00}_{S,\,q}(k\gg k_0)=\frac{2ig^2n_f}{\pi k}\int_{k/2}^\infty\,dq\,q^2\left(1-\frac{k^2}{4q^2}\right)
\,n_\mathrm{F}(q)[1-n_\mathrm{F}(q)]\,,
\label{pi00fermion}
\end{equation}
where, here and in the following, we neglect contributions of order $k_0^2/k^2$ or smaller; 
$n_f$ is the number of light quarks in the self-energy loop. 
Noticing that $-T dn_\mathrm{F}(q)/dq = n_\mathrm{F}(q)[1-n_\mathrm{F}(q))]$, eq.~\eqref{pi00fermion} can be rewritten as
\begin{equation}
\Pi^{00}_{S,\,q}(k\gg k_0)=\frac{4ig^2n_f T}{\pi k}\int_{k/2}^\infty\,dq\,q\,n_\mathrm{F}(q)\,,
\label{intbyparts}
\end{equation}
which agrees with the expression of $\Pi^{00}_{S,\,q}(k\gg k_0)$ that follows from eq.~(45) of~\cite{Brambilla:2008cx}. 
Si\-milarly, the one-loop gluon contribution to the longitudinal polarization tensor in Coulomb gauge reads
\begin{equation}
\Pi^{00}_{S,\,g}(k\gg k_0)=\frac{2 ig^2 N_c}{\pi k}\int_{k/2}^\infty\,dq\,q^2
\left(1-\frac{k^2}{2q^2}+\frac{k^4}{8q^4}\right)\,n_\mathrm{B}(q)[1+n_\mathrm{B}(q)]\,,
\label{pi00gluon}
\end{equation}
where $\nc$ is the number of colours. 
Again, noticing that $-T dn_\mathrm{B}(q)/dq = n_\mathrm{B}(q)[1+n_\mathrm{B}(q)]$, 
eq.~\eqref{pi00gluon} can be rewritten as 
\begin{equation}
\Pi^{00}_{S,\,g}(k\gg k_0)=\frac{4 ig^2 N_c T}{\pi k}\left[\frac{k^2}{8}n_\mathrm{B}(k/2)
+  \int_{k/2}^\infty\,dq\,q \left(1-\frac{k^4}{8q^4}\right)\,n_\mathrm{B}(q)\right]\,,
\label{intbypartsg}
\end{equation}
which also agrees with the expression of $\Pi^{00}_{S,\,g}(k\gg k_0)$ that follows from eq.~(45) of~\cite{Brambilla:2008cx}. 

In section~\ref{sec_eggmd} we will also need the symmetric transverse self-energy in the hard thermal loop (HTL) 
approximation $k_0, k\ll T$ \cite{Braaten:1989mz} . It reads
\begin{equation}
\label{symmtransverseHTL}
\Pi^{T}_S(k_0,k)=\frac{i g^2}{\pi k}\theta(k^2-k_0^2)\left(\frac{k_0^2}{k^2}-1\right)\int_0^\infty dq q^2\left(
\nc\,n_\mathrm{B}(q)[1+n_\mathrm{B}(q)]+n_f\,n_\mathrm{F}(q)[1-n_\mathrm{F}(q)]\right),
\end{equation}
where $\Pi^T\equiv(\delta^{ij}-\hat{k}^i\hat{k}^j)\Pi^{ij}/2$ and we have neglected higher-order corrections 
in $k_0/T$.

Equations \eqref{pi00fermion}, \eqref{pi00gluon} and \eqref{symmtransverseHTL} suggest the following form 
for the parton-scattering dissociation width:
\begin{equation}
\Gamma_{\rm HQ}=\sum_p\int_{q_\mathrm{min}}\frac{d^3q}{(2\pi)^3}\,f_p(q)\,\left[1\pm f_p(q)\right]\,\sigma_p^{\rm HQ}(q)\,,
\label{eq:sec2}
\end{equation} 
where the plus sign applies when the parton is a boson and the minus sign when the parton is a fermion.
This expression incorporates the quantum-statistical effects both of Pauli blocking on the light-quark final states 
and of Bose enhancement on the gluon final states.
We have set $q^0_\mathrm{in}=q^0_\mathrm{out}=q$, which is a good approximation as
long as $T \gg E$, since the incoming parton is on shell (hence $q^0_\mathrm{in}=q$) 
and its momentum is of the order of the temperature, while the transferred energy is of the order of the binding energy.
Although eqs.~\eqref{intbyparts} and \eqref{intbypartsg} seem to allow for a  dissociation width 
of the form (\ref{eq:facw}), this is actually not the case if $\sigma_p^{\rm HQ}$ has to be understood as a cross section.
In fact the quantity convoluted with the distribution functions, 
when using \eqref{intbyparts} and \eqref{intbypartsg}, cannot be interpreted as a parton-heavy-quarkonium cross section 
as it does not follow from applying the optical theorem to a quarkonium-quarkonium amplitude.
Our conclusion is therefore that eq.~(\ref{eq:facw}) in its common interpretation is not justified by QCD at finite temperature. 
QCD at finite temperature suggests instead formula~(\ref{eq:sec2}) or its generalization 
for the case $q^0_{\mathrm{in}}\ne q^0_\mathrm{out}$.
In the rest of the paper, we will explicitly derive eq.~(\ref{eq:sec2}) at leading order  
for a wide range of temperatures.

The momentum $q_\mathrm{min}$ is equal to the absolute value of the quarkonium 
binding energy. As long as $T \gg E$, which will be the case for all thermal 
regimes discussed in this paper, we can set $q_\mathrm{min}=0$ in the convolution 
integral \eqref{eq:sec2}. Corrections in $q_\mathrm{min}/T$ are  suppressed.

As a final comment, we remark that eq.~\eqref{eq:sec2} only
holds in a leading-order picture, where one light parton with momentum
of order $T$ scatters off the bound state. At higher order, when more partons appear in the 
initial or final states, or when some of them have momenta of order $gT$, 
eq.~\eqref{eq:sec2} is no longer valid.

\section{General considerations on the EFT approach}
\label{sec_EFT} 
Before calculating the dissociation width and cross section 
in an EFT framework, we summarize here some general aspects of this framework.
As mentioned in the introduction, this approach is based on the
hierarchies of non-relativistic  and thermal scales typical of
quarkonium in a quark-gluon plasma.  
For definiteness, we will assume that the non-relativistic scales, 
the temperature and the Debye mass are larger than the typical hadronic scale $\lqcd$, 
which justifies a perturbative treatment for all of them; this also implies that $v\sim\als$.
A system whose energy scales may possibly satisfy this assumption 
is the bottomonium ground state at LHC~\cite{Vairo:2010bm}. 

In the following, we will consider several possible temperature regimes at weak coupling. 
We will proceed integrating out from thermal QCD all scales larger than $E\sim mv^2$. 
The ultimate EFT that describes $Q\overline{Q}$ pairs with momentum of order $mv$ and energy 
of order $mv^2$ interacting with gluons and light quarks of energy and momentum 
of order $mv^2$ or smaller has the form of pNRQCD. If the temperature is larger 
than $mv^2$, the matching coefficients of pNRQCD will depend on the temperature.
At the accuracy we will need it in the following, the pNRQCD Lagrangian 
at weak coupling reads~\cite{Pineda:1997bj,Brambilla:1999xf}
\begin{eqnarray}
{\cal L}_{\textrm{pNRQCD}} &=& 
{\cal L}_{\textrm{light}}
+ \int d^3r \; {\rm Tr} \,  
\Bigl\{ {\rm S}^\dagger \left[ i\partial_0 - h_s \right] {\rm S} 
+ {\rm O}^\dagger \left[ iD_0 -h_o \right] {\rm O} 
\nonumber \\
&& \hspace{1.8cm}
+ 
{\rm O}^\dagger \br \cdot g\be \,{\rm S} + 
{\rm S}^\dagger \br \cdot g\be \,{\rm O}
+ \frac{1}{2} {\rm O}^\dagger \left\{ \br\cdot g\be \,, {\rm O}\right\} 
\Bigr\} \,,
\label{pNRQCD}	
\end{eqnarray}
where ${\cal L}_{\textrm{light}}$ is the part of the Lagrangian that describes the propagation of light quarks and gluons, 
$\mathrm{S}=S\,\mathbf{1}_c/\sqrt{N_c}$ and $\mathrm{O}=\sqrt{2} O^a\,T^a$ are 
the $Q\overline{Q}$ colour-singlet and colour-octet fields respectively, 
$\be$ is the chromoelectric field and $iD_0 \mathrm{O} =i\partial_0 \mathrm{O} - gA_0 \mathrm{O} + \mathrm{O} gA_0$. 
The trace is over colour and spin indices. Gluon fields depend only on the centre-of-mass coordinate and on time.
In \eqref{pNRQCD} we have neglected irrelevant operators of order $r^2$, $1/m$ or smaller; 
we have also neglected quantum corrections to the matching coefficients of the dipole operators, which 
are of order $\als$ or smaller, and beyond our accuracy.
In the centre-of-mass frame, the singlet and octet Hamiltonians have the form ($\bp\equiv-i\nabla_\br$): 
\begin{equation}
h_{s,o}=\frac{\bp^2}{m} +V^{(0)}_{s,o} +\frac{V^{(1)}_{s,o}}{m}+\frac{V^{(2)}_{s,o}}{m^2}+\ldots\,,
\label{sinoctham}
\end{equation}
where the dots stand for higher-order terms in the $1/m$ expansion.
The first two terms in the right-hand side, which are the kinetic energy 
and the static potential respectively, constitute the leading-order Hamiltonian.

We refer to~\cite{Brambilla:2008cx,Ghiglieri:2012rp} for details on
 pNRQCD in the context of the real-time formalism of
thermal field theory. We only mention that the ``2'' $Q\overline{Q}$ fields decouple, 
hence all singlet and octet  $Q\overline{Q}$ fields appearing in pNRQCD amplitudes 
have to be understood as  ``1'' fields.

\section{The $T\gg mv\sim m_D$ case}
\label{sec_tggr} 
In the same framework that we adopt here, the case $T\gg mv\sim m_D$ was studied 
in section VI of~\cite{Brambilla:2008cx} and in section V~B of~\cite{Escobedo:2008sy}. 
Since $m\gg T$, we start by integrating out from QCD the mass scale, obtaining NRQCD. 
Then we integrate out the scale $T$ from NRQCD to arrive at a version of NRQCD that is modified by the temperature.
In particular, the gauge and light-quark degrees of freedom are described
by the hard thermal loop Lagrangian~\cite{Braaten:1991gm,Frenkel:1989br}, 
whereas the heavy-quark sector is not modified at leading order.\footnote{
This thermal version of NRQCD was called NRQCD${}_\mathrm{HTL}$ in~\cite{Vairo:2009ih,Ghiglieri:2012rp}.
In the abelian case the corresponding EFT was named NRQED$_T$ in~\cite{Escobedo:2010tu}.} 
The next step consists in integrating out also the scales $mv$ and $m_D$ 
to obtain a version of pNRQCD specific for the hierarchy $T\gg mv\sim m_D$.\footnote{
\label{foot_pnrqcdhtl}
This version of pNRQCD was called pNRQCD${}_\mathrm{HTL}$ in~\cite{Vairo:2009ih}.
In the abelian case the corresponding EFT was named pNRQED$_T$ in~\cite{Escobedo:2010tu}. 
In~\cite{Ghiglieri:2012rp} it was instead named pNRQCD${}_{m_D}$. 
We also remark that in the literature there exist two versions of pNRQCD${}_\mathrm{HTL}$, 
one for $T\gg mv$, corresponding to the present case, and one for $mv\gg T$, 
which was studied in detail in~\cite{Brambilla:2010vq,Brambilla:2011mk}.} 
This version of pNRQCD corresponds to the Lagrangian~\eqref{pNRQCD}, with ${\cal L}_{\textrm{light}}$ 
given by the HTL Lagrangian and with thermally modified $Q\overline{Q}$ potentials. 
In particular, $V_s^{(0)}$ is at leading order the potential computed in~\cite{Laine:2006ns}; it reads 
\begin{equation}
V^{(0)}_s(r) =  -C_F\,\als \left( \frac{e^{-m_Dr}}{r} + m_D \right) 
+ i2C_F\,\als\, T\,\int_0^\infty dt \,\left(\frac{\sin(m_Dr\,t)}{m_Dr\,t}-1\right)\frac{t}{(t^2+1)^2}\,,
\label{Vsrsimmd}
\end{equation}
where $C_F = (N_c^2-1)/(2N_c)$ and 
\begin{equation}
m_D^2=\frac{g^2T^2}{3}\left(N_c+\frac{n_f}{2}\right).
\label{defmd}
\end{equation}
The imaginary part of eq.~\eqref{Vsrsimmd} describes precisely the physics of dissociation  
by inelastic parton scattering. 
Corrections of higher order in $1/m$ and $g$ are beyond the accuracy of this paper. 

At leading order in the multipole expansion, the equation of motion for the singlet field
resulting from the pNRQCD Lagrangian is a Schr\"{o}dinger equation with the potential, 
provided by eq.~\eqref{Vsrsimmd}, consisting of a real and an imaginary part. 
As discussed in~\cite{Brambilla:2008cx}, if $mv\sim m_D$ the latter is larger 
than the former by a factor of $T/m_D$ and the bound state can be considered dissociated. 
On the other hand, if $mv$ is sufficiently larger than $m_D$,
corresponding to the situation $T\gg mv \gg m_D$, then both the imaginary part 
and the screening are perturbations of the Coulomb potential~\cite{Escobedo:2008sy,Brambilla:2008cx}. 
The temperature at which the real and imaginary parts become of
the same size can be defined as the \emph{dissociation temperature} $T_d$. 
One then has $T_d\sim m g^{4/3}$~\cite{Escobedo:2008sy,Laine:2008cf} 
(see also~\cite{Escobedo:2010tu} for numerical estimates of the $\Upsilon(1S)$ dissociation 
temperature).

We derive now the cross section and decay width from the potential~\eqref{Vsrsimmd} under the 
assumption that the real part of the potential is larger than its imaginary part.
At leading order, the decay width is given by
\begin{equation}
\Gamma_{nl}= - \langle n,l|2\,\mathrm{Im}\,V^{(0)}_s(r)|n,l\rangle\,,
\label{defwidth}
\end{equation}
where $|n,l\rangle$ is an eigenstate of ${\bf p}^2/m + \mathrm{Re}\, V^{(0)}_s(r)$, and $n$, $l$ are 
the principal and orbital angular momentum quantum numbers identifying the quarkonium.
From eq.~\eqref{Vsrsimmd}, it follows that 
\begin{equation}
-2\,\mathrm{Im}\,V^{(0)}_s(r)= \frac{g^2 T \cf m_D^2}{\pi}\int_0^\infty\frac{dt\,t}{(t^2+m_D^2)^2}\left(1-\frac{\sin(tr)}{tr}\right).
\label{rewriteimvs}
\end{equation} 
We observe that, given eq.~\eqref{defmd}, the imaginary part of the potential can be 
separated in a part coming from the scattering with light quarks and in a part coming from the scattering with gluons. 
More in detail, following the arguments of section~\ref{sec:gen}, the imaginary part originates
from the symmetric part of the longitudinal propagator taken in the HTL limit, $k_0, k \ll T$. 
The longitudinal propagator is given by eq.~\eqref{deltaspi}. The longitudinal polarization tensor, $\Pi^{00}_S(k_0,k)$,  
in the HTL limit follows from eqs.~\eqref{pi00fermion} and \eqref{pi00gluon} expanded for $k_0, k \ll T$.
For the quark contribution we have
\begin{equation}
-\frac{ik}{2\pi T}\Pi^{00}_{S,\,q}(k_0=0,k\ll T)=\frac{2 g^2 n_f}{T} 
\int\frac{d^3q}{(2\pi)^3}n_\mathrm{F}(q)\left[1-n_\mathrm{F}(q)\right] = m_D^2|_\mathrm{quark}  \,,
\label{eq:rel4q}
\end{equation}
where $m_D^2|_\mathrm{quark}= {g^2n_fT^2}/{6}$. For the gluon contribution we have 
\begin{equation}
-\frac{ik}{2\pi T}\Pi^{00}_{S,\,g}(k_0=0,k\ll T)=\frac{2 g^2 \nc}{T} 
\int\frac{d^3q}{(2\pi)^3}n_\mathrm{B}(q)\left[1+n_\mathrm{B}(q)\right] = m_D^2|_\mathrm{gluon}\,,
\label{eq:rel4g}
\end{equation}
where $m_D^2|_\mathrm{gluon}= {g^2\nc T^2}/{3}$.
We can then rewrite the width as
\begin{equation}
\Gamma_{nl}=\int\frac{d^3q}{(2\pi)^3}\bigg[n_\mathrm{F}(q)(1-n_\mathrm{F}(q))\langle n,l|\Sigma_q(r,q)| n,l\rangle
+n_\mathrm{B}(q)(1+n_\mathrm{B}(q))\langle n,l|\Sigma_g(r,q)| n,l\rangle\bigg],
\label{defsigma}
\end{equation} 
where 
\begin{equation}
\Sigma_q(q,r)=32\pi C_Fn_f\als^2\int_0^\infty\frac{dt\, t}{(t^2+m_D^2)^2}\left(1-\frac{\sin(tr)}{tr}\right),
\label{sigmaq}
\end{equation}
and
\begin{equation}
\Sigma_g(q,r)=32\pi C_FN_c\als^2\int_0^\infty\frac{dt\, t}{(t^2+m_D^2)^2}\left(1-\frac{\sin(tr)}{tr}\right).
\label{sigmag}
\end{equation} 
Finally, we can identify 
\begin{equation}
\sigma_p^{nl}(q) = \langle n,l|\Sigma_{p}(r,q)|n,l\rangle\,,
\label{crossTmv}
\end{equation}
with the cross section of a quarkonium state (with quantum numbers $n,l$) with a parton $p = q,g$ in the medium, 
and arrive at the formula \eqref{eq:sec2}. 
Note that the gluon- and quark-induced cross sections differ only in the colour structure. 

We conclude this section with some comments about the cross sections.
First, we note that $\Sigma_p(q,r)$ and hence the cross sections $\sigma_p^{nl}(q)$
do not depend on the parton momentum, $q\sim T$. This holds at leading order for momenta such that $m \gg q \gg mv$. 
As can be seen from eqs.~\eqref{pi00fermion} and \eqref{pi00gluon}, it is a consequence 
of assuming the temperature to be much larger than the other scales, $mv$, $m_D$ and $E$. 
We also remark that the cross sections depend on the
temperature only through the Debye mass, $m_D$. Hence, in contrast to what happens 
for the gluo-dissociation cross section~\cite{Brambilla:2011sg}, 
we cannot relate the cross section for inelastic parton scattering,
 not even at leading order, with a zero temperature process. 
The underlying reason is the infrared sensitivity of the cross section at the momentum scale $mv$, 
which can be seen by putting $m_D=0$ in eqs.~\eqref{sigmaq} and \eqref{sigmag}. 
This infrared sensitivity is cured by the HTL resummation at the scale $m_D$, 
as it will become more apparent in the next section.

\subsection{The $mv \gg m_D$ case}
\label{submvggmD}
We consider now the cross section and decay width of a quarkonium state in the special case $mv\gg m_D$. 
Under this condition the state is Coulombic, i.e. $\mathrm{Re}\, V^{(0)}_s(r) \approx -C_F\als/r$. 
This case may be possibly realized only for a quarkonium $1S$ state~\cite{Vairo:2010bm}, 
which we are going to consider in the following.
The $1S$ wave function is given by $\langle\br\vert1S\rangle=1/(\sqrt{\pi}a_0^{3/2})\exp(-r/a_0)$, 
where $a_0=2/(m \cf \als )$ is the Bohr radius. 
The evaluation of the dissociation cross section follows then easily from 
\begin{equation}
\langle 1S|\frac{\sin(tr)}{tr}|1S\rangle=\frac{16}{(t^2a_0^2+4)^2}\,.
\label{eq:ref1}
\end{equation}
In particular, the light-quark cross section becomes
\begin{equation}
\sigma^{1S}_q(q) = \langle 1S|\Sigma_q(r,q)| 1S\rangle = 8\pi C_Fn_f\,\als^2a_0^2f(m_Da_0)\,,
\label{quarkrsimmd}
\end{equation}
where
\begin{equation}
f(x)=\frac{2}{x^2}\left[1-4\frac{x^4-16+8x^2\ln\left({4}/{x^2}\right)}{(x^2-4)^3}\right],
\label{frsimmd}
\end{equation}
and a very similar formula holds for the gluon cross section:
\begin{equation}
\sigma^{1S}_g(q) = \langle 1S|\Sigma_g(r,q)| 1S\rangle = 8\pi C_F\nc\,\als^2a_0^2f(m_Da_0)\,.
\label{gluonrsimmd}
\end{equation}
It is convenient to define the constants
\begin{equation}
\sigma_{cq}\equiv8\pi C_Fn_f\,\als^2\,a_0^2\,,
\label{eq:defscq}
\end{equation}
and
\begin{equation}
\sigma_{cg}\equiv8\pi C_F\nc\,\als^2\,a_0^2\,,
\label{eq:defscg}
\end{equation}
so that $\sigma^{1S}_{p}=\sigma_{cp}f(m_Da_0)$, with $p = g,q$.
Plugging eqs.~\eqref{quarkrsimmd} and \eqref{gluonrsimmd} into eq.~\eqref{defsigma}, we obtain the width, which reads
\begin{equation}
\Gamma_{1S}=2C_F\als T\,\left[1-4\frac{(m_Da_0)^4-16+8(m_Da_0)^2\ln\left({4}/{(m_Da_0)^2}\right)}{((m_Da_0)^2-4)^3}\right]. 
\label{widthrsimmd}
\end{equation}
The width in this regime was already obtained in eq.~(1.7) of~\cite{Dumitru:2010id}.
Our result appears to be larger by a factor of 2 due to the fact that 
in~\cite{Dumitru:2010id} the width is defined as one half of ours.
Under the assumption $mv\gg m_D$ we can further expand $f(m_Da_0)$ for $m_Da_0\ll 1$, obtaining for the cross section
\begin{equation}
\sigma^{1S}_p(q)=\sigma_{cp}\left(\ln\frac{4}{m_D^2a_0^2}-\frac{3}{2}\right),\quad {\rm with}\; p=q,\,g\,,
\label{crossrggmd}
\end{equation}
and for the decay width
\begin{equation}
\Gamma_{1S}=C_F\als T\, m_D^2a_0^2\left(\ln\frac{4}{m_D^2a_0^2}-\frac{3}{2}\right). 
\label{widthrsggmd}
\end{equation}

\begin{figure}[ht]
\begin{center}
\includegraphics[width=14cm]{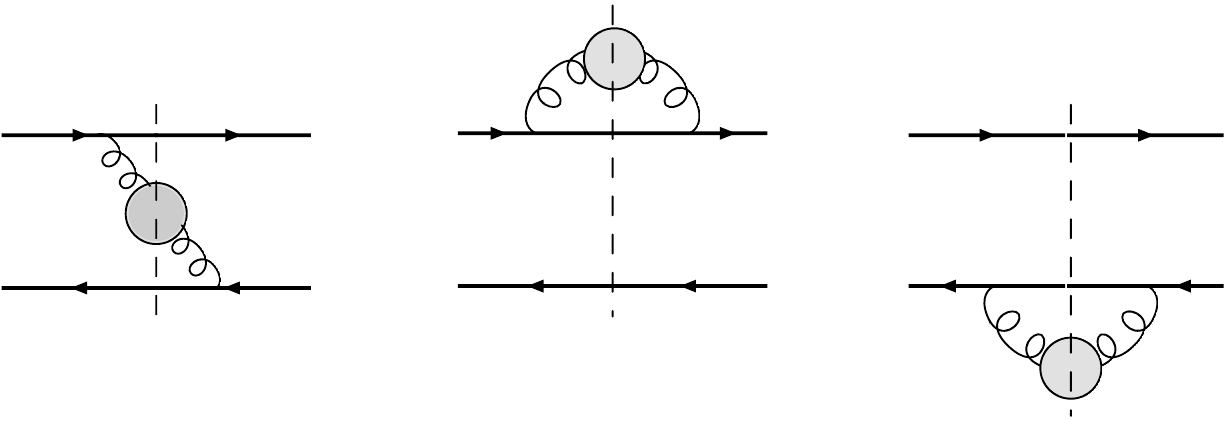}
\end{center}
\caption{Cut diagrams in NRQCD. The momentum of the cut partons is of order $T$. 
Dashed lines stand for the cuts, thick lines with arrows for heavy quarks and antiquarks, 
curly lines for gluons and grey blobs for either light-quark or gluon loops.}
\label{fig:selfcut1}
\end{figure}

\begin{figure}[ht]
\begin{center}
\includegraphics[width=5cm]{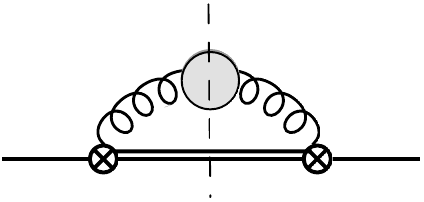}
\end{center}
\caption{Cut diagram in pNRQCD. The momentum of the cut partons is of order $T$. 
The single line stands for a $Q\overline{Q}$ pair in a colour-singlet configuration,
the double line for a $Q\overline{Q}$ pair in a colour-octet configuration, 
the circle with a cross for a chromoelectric dipole vertex, 
the curly line for a HTL gluon and the grey blob for either a light-quark or a gluon loop.
}
\label{fig:selfcut2}
\end{figure}

An alternative derivation of the cross section and decay width given in 
eqs.~\eqref{crossrggmd} and \eqref{widthrsggmd} consists in assuming right from the start that
$T\gg mv\gg m_D$ and in evaluating the potential through the hierarchy of EFTs
introduced in section VI~E of~\cite{Brambilla:2008cx}. It was shown there that the potential, 
as defined in the ultimate EFT, receives a contribution from the scale $mv$ and one from the scale $m_D$. 
The leading-order real part of the static potential is the Coulomb potential, 
which comes from the scale $mv$. For what concerns the imaginary part of the static potential, 
the contribution from the scale $mv$ is infrared divergent and 
originates from the cut diagrams in figure~\ref{fig:selfcut1} when the momentum flowing 
in the gluon is of order $mv$. It reads in dimensional regularization ($D$ is the number of spacetime dimensions)
\begin{equation}
\mathrm{Im}\,V^{mv}(r)=\frac{\cf}{6}\als r^2 T m_D^2\left(\frac{2}{4-D}+\gamma_E+\ln\pi+\ln(r\,\mu)^2-1\right),
\label{1/rcontribrggmd}
\end{equation} 
where $\gamma_E$ is the Euler--Mascheroni constant and $\mu$ is the subtraction scale. 
The contribution from the scale $m_D$ arises from the cut diagram in figure~\ref{fig:selfcut2},  
where the displayed gluon stands for a HTL-resummed gluon carrying a momentum of order $m_D$.
The gluon interacts with the $Q\overline{Q}$ field through the chromoelectric dipole 
interactions induced by the (second line of the) Lagrangian~\eqref{pNRQCD}. 
The diagram was evaluated in eq.~(87) of~\cite{Brambilla:2008cx} and found to be ultraviolet divergent. It gives
\begin{eqnarray} 
\mathrm{Im}\,V^{m_D}(r) &=&-\frac{\cf}{6}\als r^2 T m_D^2\left(\frac{2}{4-D}-\gamma_E+\ln\pi+\ln\frac{\mu^2}{m_D^2}+\frac53\right)\,,
\label{mdcontribrggmd}
\end{eqnarray}
which holds at leading order in an expansion in $E/m_D$.\footnote{
The condition $T \gg mv$ implies $m_D \gg E$ for a Coulombic bound state in a weakly-coupled plasma.
}
At this order the octet propagator in figure~\ref{fig:selfcut2} can be taken 
to be $1/(-k_0+i\epsilon)$, where $k_\mu$ is the gluon momentum; this means that 
the rescattering of the unbound final-state heavy quarks can be neglected.
Summing the two contributions the divergence and related $\mu$ dependence cancel and one obtains
\begin{equation}
\mathrm{Im}\,V^{mv\gg m_D}(r) \equiv \mathrm{Im}\,V^{mv}(r) + \mathrm{Im}\,V^{m_D}(r) 
= \frac{\cf}{6}\als r^2 T m_D^2\left(2\gamma_E+\ln(r\,m_D)^2-\frac83\right).
\label{impotrggmd}
\end{equation}
Using 
\begin{equation}
\langle 1S|r^2|1S\rangle=3a_0^2\,,
\label{eq:ref2}
\end{equation}
and
\begin{equation}
\langle 1S|r^2\ln\left(\frac{r}{a_0}\right)|1S\rangle=3a_0^2\left(\frac{25}{12}-\gamma_E-\ln 2\right)\,,
\label{eq:ref3}
\end{equation} 
from \eqref{defwidth} the thermal width \eqref{widthrsggmd} and the dissociation cross section \eqref{crossrggmd} follow. 

\begin{figure}[ht]
\begin{center}
\includegraphics[width=12cm]{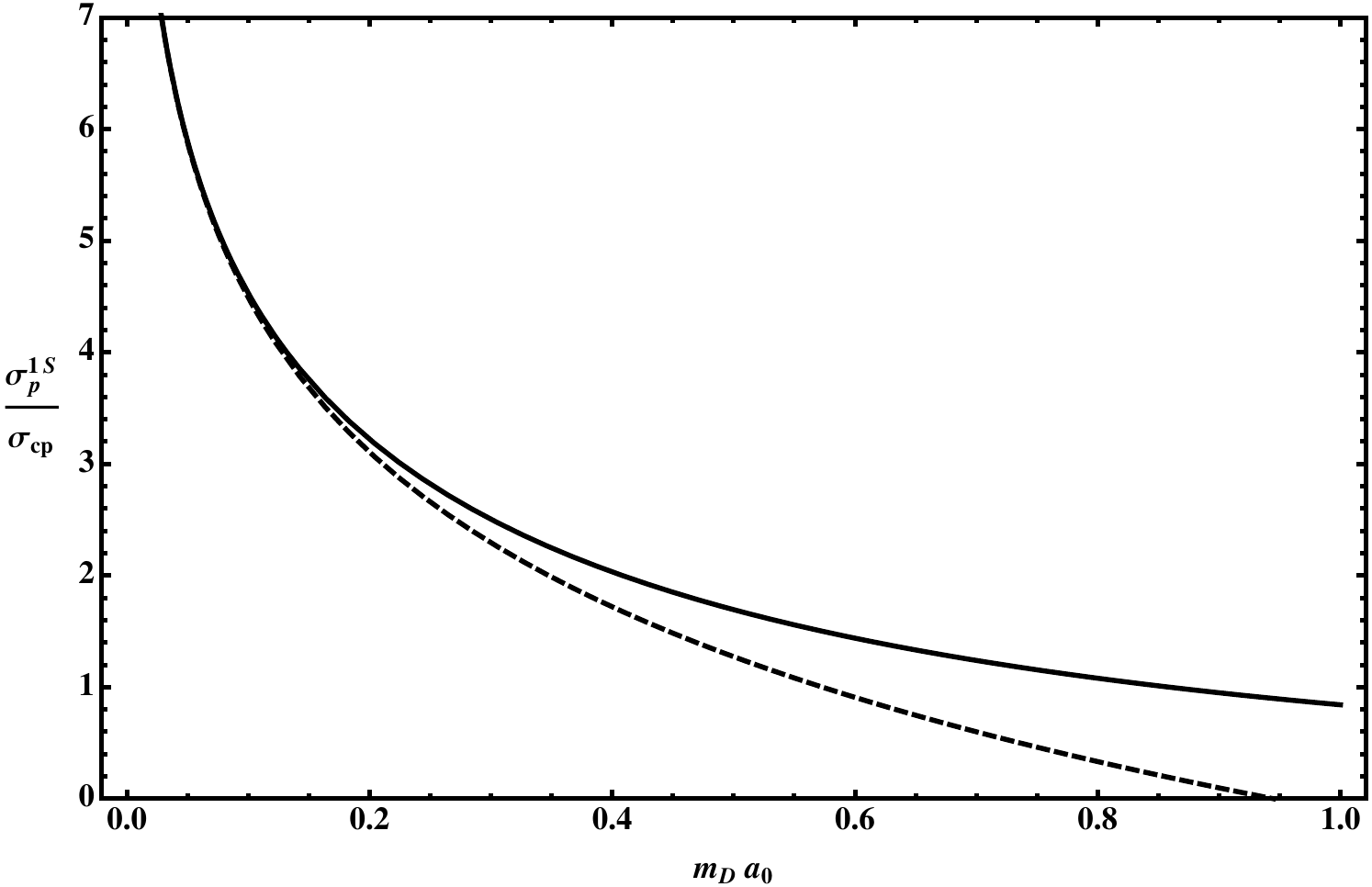}
\end{center}
\caption{The rescaled cross sections $\sigma^{1S}_p/\sigma_{cp}$ as a
  function of $m_Da_0$. The continuous line corresponds to
  $f(m_Da_0)$, as defined in eq.~\eqref{frsimmd}, whereas the dashed
  line follows from eq.~\eqref{crossrggmd} and is thus the expansion of
  $f(m_Da_0)$ for $m_Da_0 \ll 1$. 
}
\label{tggr}
\end{figure}  

In figure~\ref{tggr} we plot the rescaled cross section $\sigma^{1S}_p/\sigma_{cp}$, 
obtained in eqs.~\eqref{quarkrsimmd} and \eqref{gluonrsimmd}, as well as its
expansion for $m_Da_0\ll1$, obtained in eq.~\eqref{crossrggmd}. 
They are the continuous and dashed curve respectively. 
The two curves overlap up to $m_Da_0 \approx 0.2$. 
We note that the curves have physical meaning only for values of $m_Da_0$ significantly smaller than~$1$.
The reasons are that the dashed curve has been obtained as the leading term of an expansion in $m_Da_0$, 
and the continuous curve, although following from the imaginary part 
of the potential \eqref{Vsrsimmd}, which is valid also for $m_Da_0 \approx 1$, 
assumes the bound state to be Coulombic, see eq.~\eqref{eq:ref1}, 
which is instead a valid assumption only for $m_Da_0$ much smaller than $1$.

\subsection{Comparison with the literature}
As we have already mentioned,  dissociation by inelastic parton scattering was first considered 
in the context of heavy-ion collisions in~\cite{Grandchamp:2001pf}. 
There the cross section was approximated by twice the parton heavy-quark scattering cross section computed in~\cite{Combridge:1978kx},
an approximation called quasi-free approximation, complemented by the addition of momentum-independent thermal masses.  
The only dependence on the bound-state properties comes through $q_\mathrm{min}$, 
which was obtained from an in-medium binding energy calculated from a phenomenological potential model.

\begin{figure}[ht]
\begin{center}
\includegraphics[width=15cm]{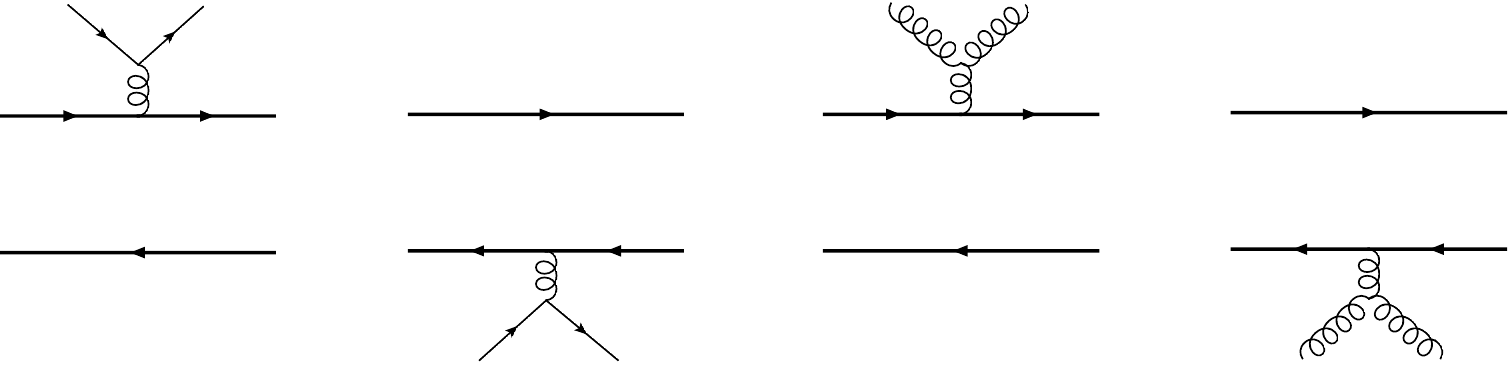}
\end{center}
\caption{Diagrams contributing to the (amplitude of the) dissociation cross sections 
  derived in section~\ref{sec_tggr} and given by eq.~\eqref{crossTmv}. 
  The thick lines represent heavy quarks and antiquarks while the thinner lines 
  represent light quarks from the medium.}
\label{fig:inter}
\end{figure}

The parton heavy-quark scattering cross section corresponds to the square of the first two diagrams in figure~\ref{fig:com}. 
Conversely, the optical theorem relates the cross sections that we have computed in section~\ref{sec_tggr}  
with the square of the sum of the diagrams shown in figure~\ref{fig:inter}. 
Each of these diagrams individually corresponds to those computed in~\cite{Combridge:1978kx}, 
however, once the square is taken, interference terms appear in the form of 
diagrams with gluons attached to different heavy-quark lines. 
These terms are the ones that are sensitive to the spatial separation between 
the heavy quark and antiquark, and ultimately to the bound state. 
The quasi-free approximation consists in neglecting these interference terms. 

\begin{figure}[ht]
\begin{center}
\includegraphics[width=12cm]{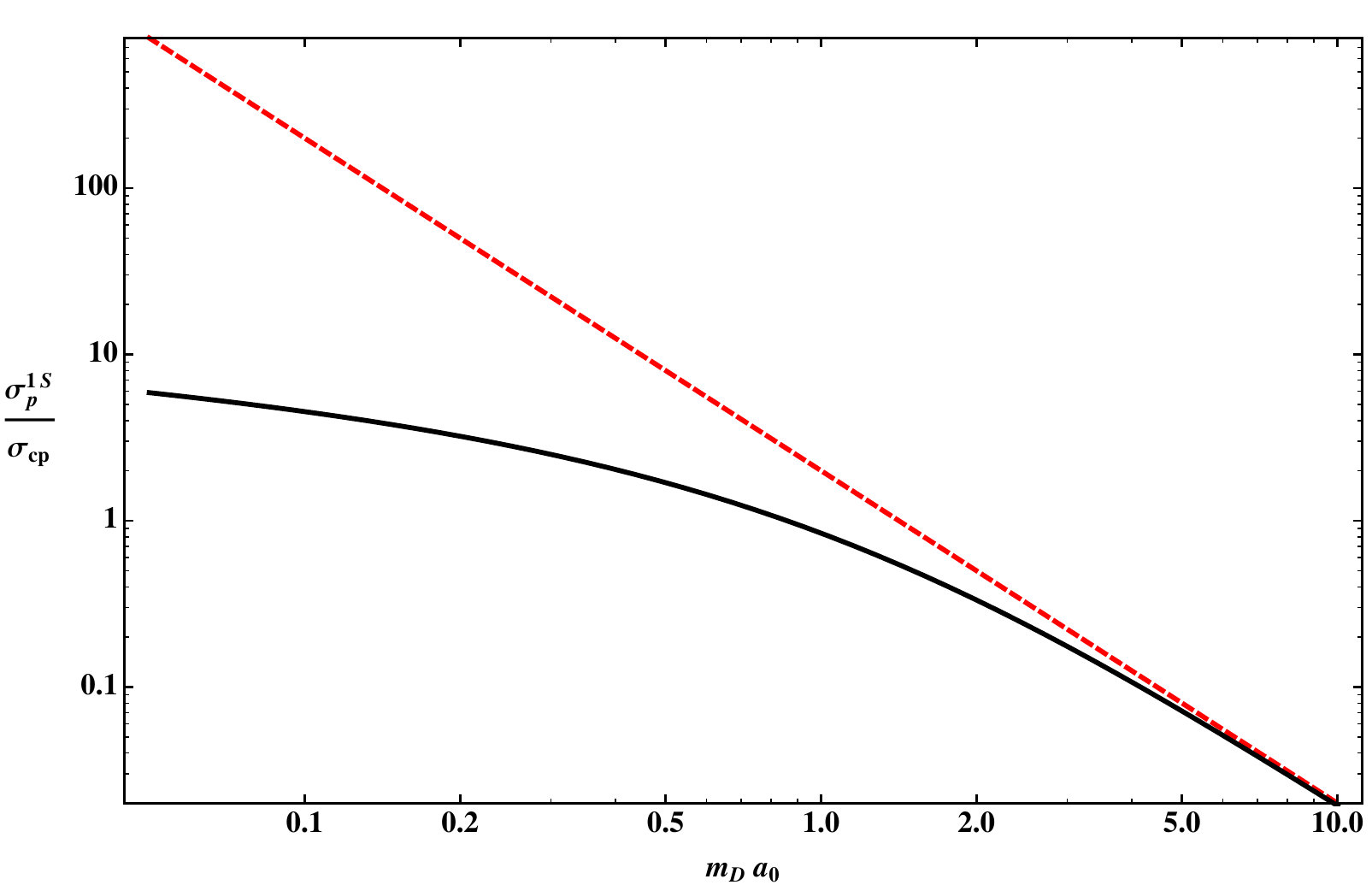}	
\end{center}
\caption{The rescaled cross section $\sigma^{1S}_p/\sigma_{cp}$ as a
  function of $m_Da_0$. The continuous line corresponds to
  $f(m_Da_0)$, as defined in eq.~\eqref{frsimmd}, whereas the red dashed
line corresponds to $2/(m_Da_0)^2$, which is the only term that survives in the quasi-free approximation.}
\label{fig_quasifreecomp}
\end{figure}

More precisely, in the right-hand side of eq.~\eqref{rewriteimvs}, the term proportional to $\sin(tr)/(tr)$ 
is the interference term while the remaining term comes from the last two cutting diagrams shown in figure~\ref{fig:selfcut1}.
These two distinct terms may be traced in the cross sections. For instance, in the Coulombic case, $mv \gg m_D$, 
we have that, in the right-hand side of eq.~\eqref{frsimmd}, the term $2/x^2$ is the contribution 
from twice the square of the parton heavy-quark scattering diagram, 
whereas the remaining terms, which are neglected in the quasi-free approximation, 
are the interference terms and give a non-trivial dependence on the wave function and thus
on the properties of the bound state. In figure~\ref{fig_quasifreecomp} we plot for comparison
$\sigma^{1S}_p/\sigma_{cp}=f(m_Da_0)$ and the quasi-free term $2/(m_Da_0)^2$. 
For small values of $m_Da_0$, where \eqref{quarkrsimmd} and \eqref{gluonrsimmd} are 
valid, the quasi-free approximation overestimates the cross section
by more than one order of magnitude. Conversely, for large values of $m_Da_0$, where the 
quasi-free approximation is sensible, the two curves overlap. In this region, however, 
we cannot treat the quarkonium as a Coulombic bound state, and, therefore, 
eqs.~\eqref{quarkrsimmd} and \eqref{gluonrsimmd} are no longer valid.

It is furthermore important to note that under the condition $mv \gg m_D$ the parton heavy-quark scattering contribution 
to the  dissociation cross section is exactly cancelled at leading order in $m_Da_0$ 
by a contribution coming from the interference terms. The remnant of this cancellation, which 
is entirely due to interference/bound-state terms, is what makes up for eq.~\eqref{crossrggmd}.
The cancellation was noticed in~\cite{Beraudo:2007ky,Brambilla:2008cx}.

In summary, the quasi-free approximation is justified when all relevant thermal scales and, 
in particular, $m_D$ are much larger than $mv$.
However, in a weak-coupling regime, the condition $m_D\gg mv$ requires a temperature $T \gg m g$,
which is parametrically larger than the dissociation temperature $T_d\sim m g^{4/3}$. 
On the other hand, whenever $m_D \ll mv$, not only the quasi-free approximation is not justified, but 
its contribution is exactly cancelled by bound-state effects.
We conclude, therefore, that, at least in a weak-coupling framework, the quasi-free approximation 
is not justified for all range of temperatures where a quarkonium can exist.

\section{The $T\sim mv \gg m_D$ case}
\label{sec_tsimr} 
The temperature region $T\sim mv \gg m_D$ was studied in detail for the case of muonic hydrogen in~\cite{Escobedo:2010tu}. 
In the muonic hydrogen case, which may be interpreted as an abelian version of heavy quarkonium,  
thermal corrections are encoded in electron loops.
In~\cite{Escobedo:2010tu} both the real and the imaginary part of the potential were computed.
Here we focus on the imaginary part of the potential, which is the quantity relevant for dissociation; 
thermal corrections are encoded in light-quark and gluon loops.

Let us briefly sketch our procedure. Since $m\gg T$ we can use NRQCD
as a starting point. Then, as $T\sim mv$, these two scales have to be integrated out at the same time. 
In doing so we go from NRQCD to a new particular thermal version of pNRQCD.
In the light-quark and gluon sector it is made of the HTL Lagrangian.
In the heavy-quark sector the Lagrangian is as in~\eqref{pNRQCD} with 
the potential of a form specific to the hierarchy $T\sim mv \gg m_D$. 
It is at leading order a Coulomb potential that receives small corrections 
with a complicated functional dependence on the temperature. 
A feature of these thermal corrections is that they are infrared divergent.
The infrared divergence that we observe in the imaginary part of the potential 
cancels against ultraviolet divergent contributions coming from the scale $m_D$, 
exactly as in the previous section. 
We discuss first light-quark and then gluon contributions.

The leading light-quark contribution to the imaginary part of the static potential 
comes from the one-loop fermion contribution to the diagrams in figure~\ref{fig:selfcut1}.
At this order only the coupling of the heavy quarks with temporal gluons is relevant.
Hence, we may follow the discussion of section~\ref{sec:gen}, and the contribution reads
\begin{eqnarray}
&&\nn
\hspace{-8mm}\mathrm{Im}\,V^T_q(r) = - \frac{1}{2}\,\mathrm{Im}\,\,\mu^{4-D}\int\frac{d^{D-1}k}{(2\pi)^{D-1}}
\left(e^{i{\bf k}\cdot{\bf r}}-1\right)\,ig^2\cf\,\frac{i \Pi^{00}_{S,\,q}(k_0=0,k)}{k^4}
\\
&& \hspace{-4mm}
=2\pi g^4C_Fn_f\int\frac{\,d^3q}{(2\pi)^3}n_\mathrm{F}(q)[1-n_\mathrm{F}(q)]\,
\mu^{4-D}\int_{2q\ge k}\frac{d^{D-1}k}{(2\pi)^{D-1}}\frac{e^{i{\bf k}\cdot{\bf r}}-1}{k^5}\left(1-\frac{k^2}{4q^2}\right),
\label{eq:imvtqtsimr}
\end{eqnarray}
where we have used the symmetric longitudinal polarization tensor given in eq.~\eqref{pi00fermion}. 
The integral in $k$ has an infrared divergence that we have regulated in dimensional regularization.
This divergence cancels against contributions coming from the scale $m_D$.
The contributions from the scale $m_D$ can be evaluated from the 
cut diagram in figure~\ref{fig:selfcut2} when the momentum flowing 
in the loop is of the order of the Debye mass.
Since $E \sim m\als^2$ is parametrically smaller than $m_D$ (exactly by one power of $g$),  
we can expand the intermediate octet propagator for $E \ll m_D$ neglecting rescattering effects 
and the result is the same as in eq.~\eqref{mdcontribrggmd}. Because that equation sums the
contribution of quarks and gluons into $m_D^2$, we need to use 
eqs.~\eqref{defmd} and \eqref{eq:rel4q} to disentangle the one from the other.
Summing the quark contribution from the scales $T\sim mv$ 
with the quark contribution from the scale $m_D$ the divergences cancel and 
the light-quark contribution to the imaginary part of the potential reads 
\begin{eqnarray}
\nn
&&\hspace{-3mm}\mathrm{Im}\,V_q^{T\sim mv}(r) \equiv \mathrm{Im}\,V_q^{T}(r) + \left.\mathrm{Im}\,V^{m_D}(r)\right\vert_{\rm quark} 
=-\frac{ g^4C_Fn_f}{\pi}\left\{\frac{r^2T^3}{144}\left(\frac83-2\gamma_E-\ln(r^2 m_D^2)\right)\right.
\\
\nn&&+\int\frac{d^3q}{(2\pi)^3}n_\mathrm{F}(q)[1-n_\mathrm{F}(q)]\left[\int_{2q}^\infty\frac{\,dt}{t^3}\left(\frac{\sin(tr)}{tr}-1\right)
\left.+\int_0^{2q}\frac{\,dt}{4tq^2}\left(\frac{\sin(tr)}{tr}-1\right)\right]\right\}.\\
&&\label{quarkpot}
\end{eqnarray}
The decay width follows from eq.~\eqref{defwidth}. It can be written in the form of eq.~\eqref{defsigma}, with
\begin{eqnarray}
\Sigma_q(r,q)&=&\frac{g^4 C_F n_f \,r^2}{3\pi}\left[\frac{4}{3}-\gamma_E-\ln(r m_D)+\frac{6}{r^2}
\int_{2q}^\infty\frac{\,dt}{t^3}\left(\frac{\sin(tr)}{tr}-1\right)\right. 
\nn \\
&&\hspace{4.5cm}\left.+\frac{3}{2q^2r^2}\int_0^{2q}\frac{\,dt}{t}\left(\frac{\sin(tr)}{tr}-1\right)\right].
\label{quarkcontribSigmatsimr}
\end{eqnarray}

\begin{figure}[ht]
\begin{center}
\includegraphics[width=12cm]{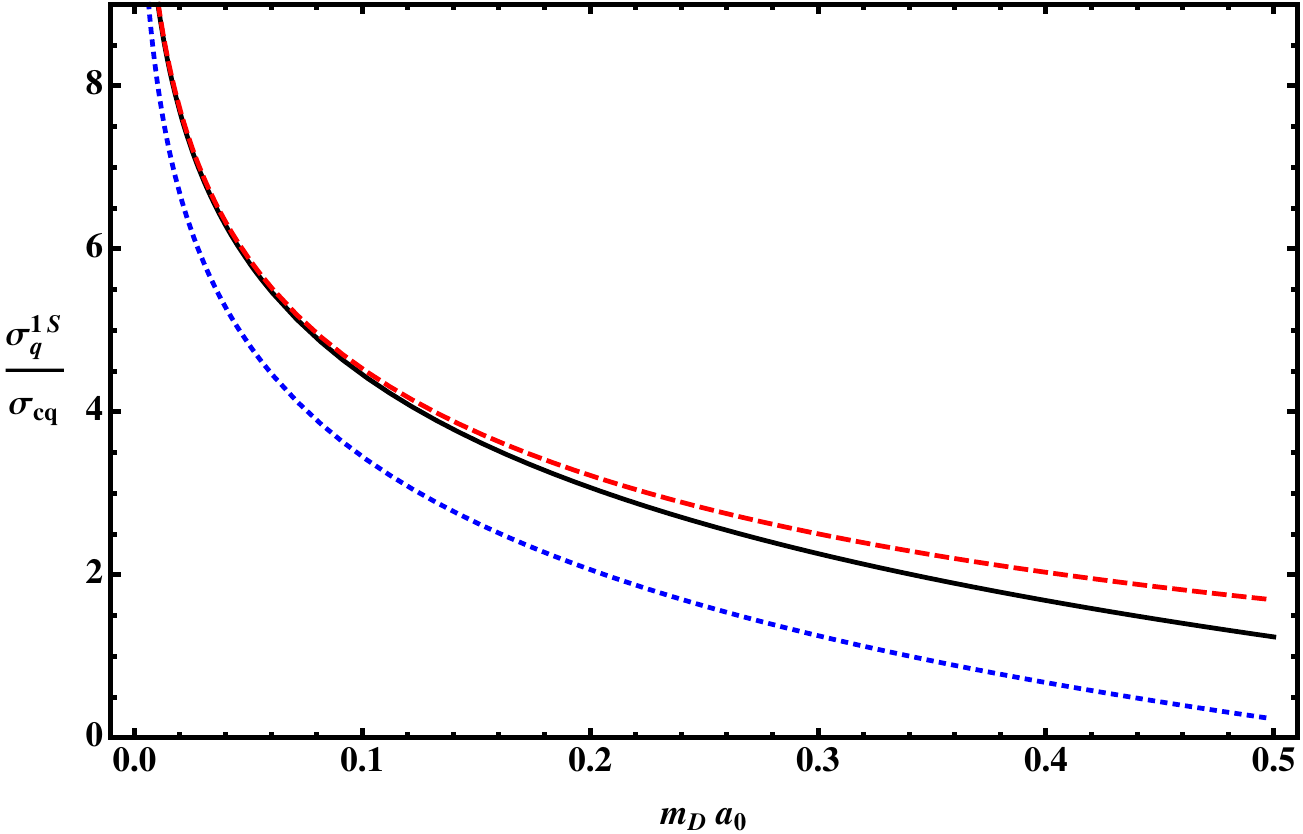}
\end{center}
\caption{Comparison between the rescaled cross sections
  $\sigma^{1S}_q/\sigma_{cq}$ for $T\gg mv\sim m_D$  and $T\sim mv\gg m_D$, 
  corresponding to eqs.~\eqref{quarkrsimmd} and~\eqref{crossquarktsimr} respectively.  
  The dashed red line is the function $f(m_Da_0)$, the black continuous line is the function
  $h_q(m_Da_0,10)$ and the dotted blue line is $h_q(m_Da_0,1)$.}
\label{fig:plottsimr}
\end{figure}

For $T\sim mv \gg m_D$, the cross section of a weakly-coupled $1S$ quarkonium state with light quarks from the medium 
is given by 
\begin{eqnarray}
\sigma^{1S}_q(q)&=&
\sigma_{cq}\, h_q(m_Da_0,qa_0)\,,
\label{crossquarktsimr}
\end{eqnarray}
where 
\begin{equation}
h_q(x,y) = -\ln\left(\frac{x^2}{4}\right)-\frac{3}{2}
+\ln\left(\frac{y^2}{1+y^2}\right)-\frac{1}{2y^2}\ln(1+y^2) \,,
\label{hq}
\end{equation}
$\sigma_{cq}$ is defined as in eq.~\eqref{eq:defscq} 
and we have made use of the matrix elements computed in eqs.~(\ref{eq:ref1}), \eqref{eq:ref2} and \eqref{eq:ref3}. 
In the previous section we derived that, in the case $T\gg mv\sim m_D$,  
$\sigma^{1S}_q(q)=\sigma_{cq}f(m_Da_0)$, 
with $f$ given in eq.~\eqref{frsimmd}. 
For  $T\sim mv\gg m_D$, and differently from the case  $T\gg mv\sim m_D$, the cross section depends on the momentum. 
However, since $T\sim mv\gg m_D$ implies that $m_Da_0\ll 1$ and $qa_0\sim 1$, 
in the limit $m_Da_0\to 0$ and $qa_0\to\infty$ the functions $f$ and $h_q$ should coincide. 
This can be seen in figure~\ref{fig:plottsimr}, where we have plotted $h_q$ as a function of $m_Da_0$ 
for $q a_0=10$ (for larger values of $qa_0$ the plot would not change significantly). 
The residual difference between $f(m_Da_0)$ and $h_q(m_Da_0,10)$ for moderate values of $m_Da_0$ 
is due to the fact that in computing $f$ we have resummed the HTL interaction while
in computing $h_q$ we have not; the difference is however a small perturbation as long as $m_Da_0 \ll 1$. 

The leading gluon contribution to the imaginary part of the static potential can be computed 
similarly to the quark contribution. From the scale $T$, we have
\begin{eqnarray}
\nn &&\hspace{-3mm} \mathrm{Im}\,V^T_g(r)= 
-\frac{1}{2}\,\mathrm{Im}\,\,\mu^{4-D} \int\frac{d^{D-1}k}{(2\pi)^{D-1}}\left(e^{i{\bf k}\cdot{\bf r}}-1\right)
\,ig^2\cf\,\frac{i \Pi^{00}_{S,\,g}(k_0=0,k)}{k^4}\\
\nn&&\hspace{2mm}
=2\pi g^4C_FN_c\int\frac{\,d^3q}{(2\pi)^3}n_\mathrm{B}(q)[1+n_\mathrm{B}(q)]\,
\mu^{4-D}\int_{2q\ge k}\frac{d^{D-1}k}{(2\pi)^{D-1}}\frac{e^{i{\bf k}\cdot{\bf r}}-1}{k^5}\left(1-\frac{k^2}{2q^2}+\frac{k^4}{8q^4}\right),
\\
\label{imvgtsimr}
\end{eqnarray}
where we have used the symmetric longitudinal polarization tensor given in eq.~\eqref{pi00gluon}.
As in the quark case, the integral in $k$ has an infrared divergence that  
cancels against the contribution coming from the scale $m_D$.
The contribution from the scale $m_D$ is given in eq.~\eqref{mdcontribrggmd}, whose  
gluonic part can be disentangled by means of eqs.~\eqref{defmd} and~\eqref{eq:rel4g}.
Summing the gluon contribution from the scales $T\sim mv$ 
with the gluon contribution from the scale $m_D$ the complete leading-order gluon contribution 
to the imaginary part of the potential reads 
\begin{eqnarray}
\nn
&& \hspace{-3mm} \mathrm{Im}\,V_g^{T\sim mv}(r) \equiv \mathrm{Im}\,V_g^{T}(r) + \left.\mathrm{Im}\,V^{m_D}(r)\right\vert_{\rm gluon} 
= -\frac{ g^4C_F\nc}{\pi}\left\{\frac{r^2T^3}{72}\left(\frac83-2\gamma_E-\ln(r^2 m_D^2)\right)\right.
\\
\nn&&+\int\frac{d^3q}{(2\pi)^3}n_\mathrm{B}(q)[1+n_\mathrm{B}(q)]
\left[\int_{2q}^\infty\frac{\,dt}{t^3}\left(\frac{\sin(tr)}{tr}-1\right)\right.
+\int_0^{2q}\frac{\,dt}{2tq^2}\left(\frac{\sin(tr)}{tr}-1\right)
\\
&&\left.\left.-\frac{1}{4q^2}\left(\frac{\sin^2(qr)}{(qr)^2}-1\right)\right]\right\}.
\label{gluonpot}
\end{eqnarray}
The decay width follows from eq.~\eqref{defwidth}. It can be written in the form of eq.~\eqref{defsigma}, with
\begin{eqnarray}
\Sigma_g(r,q)&=&
\frac{g^4 C_F N_c r^2}{3\pi}\left[\frac{4}{3}-\gamma_E-\ln(r m_D)
+\frac{6}{r^2}\int_{2q}^\infty\frac{\,dt}{t^3}\left(\frac{\sin(tr)}{tr}-1\right)\right. 
\nonumber\\
&&\hspace{2cm}
\left.+\frac{3}{q^2r^2}\int_0^{2q}\frac{\,dt}{t}\left(\frac{\sin(tr)}{tr}-1\right)-\frac{3}{2q^2r^2}\left(\frac{\sin^2(qr)}{(qr)^2}-1\right)\right].
\label{gluoncontribSigmatsimr}
\end{eqnarray}

\begin{figure}[ht]
\begin{center}
\includegraphics[width=12cm]{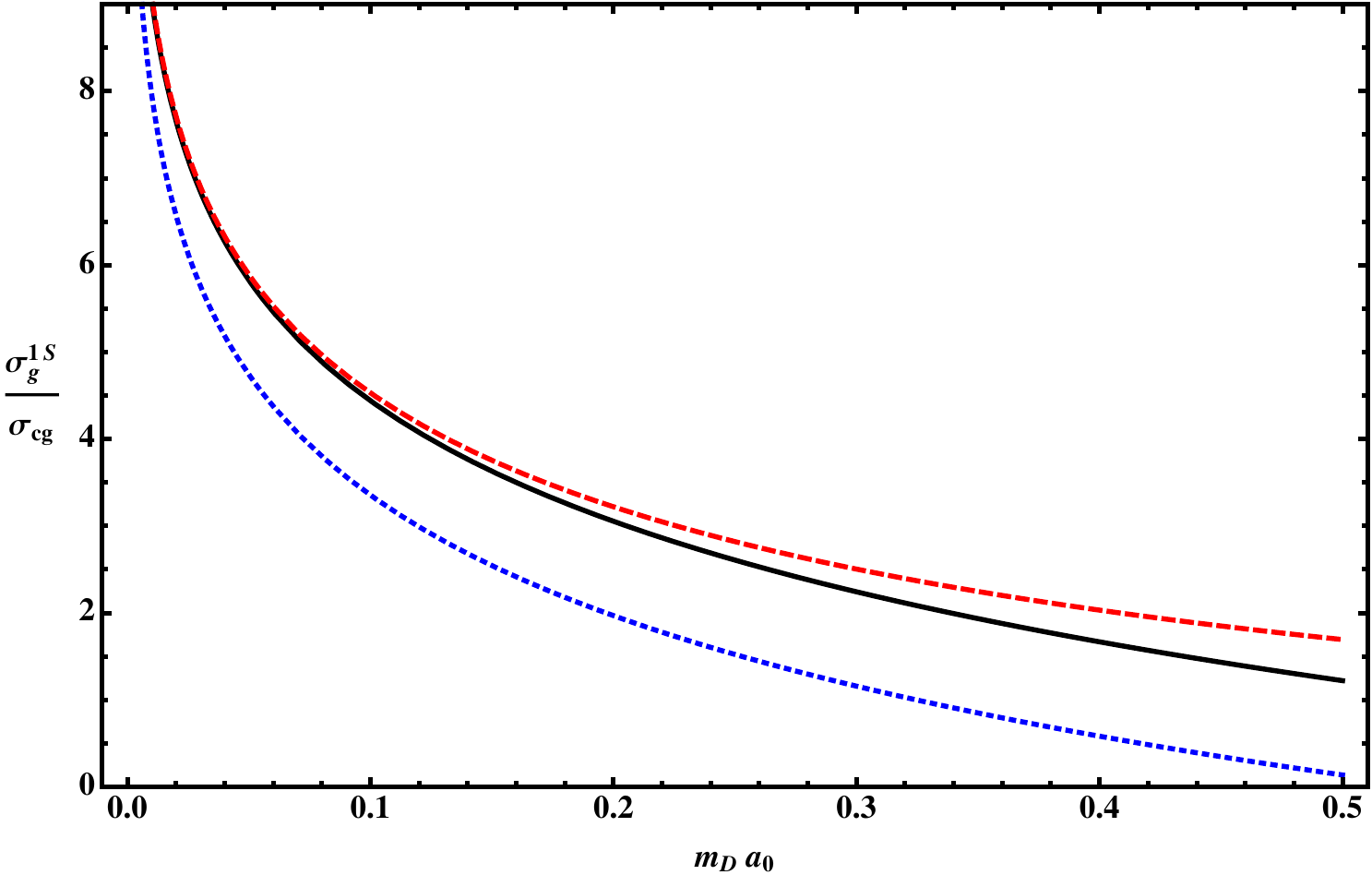}
\end{center}
\caption{
Comparison between the rescaled cross sections $\sigma^{1S}_g/\sigma_{cg}$ 
for $T\gg mv\sim m_D$  and $T\sim mv\gg m_D$, corresponding to 
eqs.~\eqref{gluonrsimmd} and~\eqref{crossgluontsimr} respectively.
The dashed red line is the function $f(m_Da_0)$, the black continuous line is the function 
$h_g(m_Da_0,10)$ and the dotted blue line is $h_g(m_Da_0,1)$.}
\label{fig:plottsimrb}
\end{figure}

For $T\sim mv \gg m_D$, the cross section of a weakly-coupled $1S$ quarkonium state with gluons from the medium 
is given by 
\begin{eqnarray}
\sigma^{1S}_g(q)&=& \sigma_{cg}\, h_g(m_Da_0,qa_0)\,,
\label{crossgluontsimr}
\end{eqnarray}
where 
\begin{equation}
h_g(x,y) = 
-\ln\left(\frac{x^2}{4}\right)-\frac{3}{2} + \frac{1}{2(1+y^2)}
+\ln\left(\frac{y^2}{1+y^2}\right) -\frac{1}{y^2}\ln(1+y^2) \,,
\label{hg}
\end{equation}
and $\sigma_{cg}$ is defined as in eq.~\eqref{eq:defscg}.
To obtain eq.~\eqref{crossgluontsimr}, 
besides eqs.~(\ref{eq:ref1}), (\ref{eq:ref2}) and (\ref{eq:ref3}), we have made use of the matrix element 
\begin{equation}
\langle 1S|\frac{\sin^2(qr)}{(qr)^2}|1S\rangle=\frac{1}{q^2a_0^2+1}.
\label{eq:refsin}
\end{equation}
In figure~\ref{fig:plottsimrb} we plot $h_g$ and $f$ as a function of $m_Da_0$, 
and show that for large values of $qa_0$ and low values of $m_Da_0$
the cross sections calculated in this section and in section~\ref{sec_tggr} overlap.

The imaginary parts of the static potential given in eqs.~\eqref{quarkpot} and~\eqref{gluonpot} 
have been calculated here for the first time. 
We observe that the imaginary part of the potential at the scale $T$, given by eqs.~\eqref{eq:imvtqtsimr} 
and~\eqref{imvgtsimr}, incorporates both the effects of the free-quark-parton scattering in the ${\bf r}$-independent 
part and of the bound state in the ${\bf r}$-dependent part. In contrast, the contribution to the 
imaginary part of the potential coming from the scale $m_D$, given by eq.~\eqref{mdcontribrggmd}, 
is just an ${\bf r}$-dependent contribution, hence it would be set to zero in the quasi-free approximation.

\section{The $mv\gg T\gg m_D\gg E$ case}
\label{sec_rggt} 
The temperature region $mv\gg T\gg m_D\gg E$ was studied in the static case in~\cite{Brambilla:2008cx}. 
Since thermal contributions associated with higher-order terms in the $1/m$ expansion turn out to be suppressed,  
at leading order we may indeed restrict ourselves to the static case. 
Our procedure goes as follows. Because $mv\gg T$ we can use the pNRQCD Lagrangian 
at $T=0$ as our starting point. Next we integrate out the scale $T$ 
to define a version of pNRQCD specific to the hierarchy $mv\gg T$ (see footnote~\ref{foot_pnrqcdhtl}). 
Its Lagrangian was derived in~\cite{Brambilla:2010vq,Brambilla:2011mk} 
and it features thermal corrections to the potentials in the heavy-quark sector 
and the HTL Lagrangian as ${\cal L}_{\rm light}$.
The correction to the colour-singlet $Q\overline{Q}$ static potential has an infrared divergence that is
compensated, as in the cases previously discussed, by an ultraviolet divergence 
coming from the scale $m_D$. After integrating out $m_D$, we obtain 
a new version of pNRQCD whose potential incorporates corrections coming from 
the Debye-mass scale. The new potential is finite and renormalization-scheme independent.
All the effects discussed in this section are specific of having an interacting 
$Q\overline{Q}$ system and would be absent in the quasi-free approximation.

The effect of the light-quark loop at the scale $T$ comes from the diagram 
in figure~\ref{fig:selfcut2} when the momentum flowing in the loop is of order $T$.
Only the longitudinal part of the gluon propagator contributes and yields~\cite{Brambilla:2008cx}
\begin{eqnarray}
\nn\mathrm{Im}\,V^T_q(r)&=& - \frac{1}{2} \, \mu^{4-D}
\int\frac{d^Dk}{(2\pi)^D}\,\mathrm{Im}\,\left[\frac{1}{-k_0+i\epsilon}
k^2 \frac{r^2}{D-1} g^2\cf 
\frac{i \Pi^{00}_{S,\,q}(k_0,k)}{k^4} \right]
\\
\nn&=&-\frac{\pi g^4C_Fn_f r^2}{D-1}
\int\frac{\,d^3q}{(2\pi)^3}n_\mathrm{F}(q)[1-n_\mathrm{F}(q)]
\,\mu^{4-D}\int_{2q\ge k}\frac{d^{D-1}k}{(2\pi)^{D-1}}\frac{1}{k^3}\left(1-\frac{k^2}{4q^2}\right),\\
&&\label{eq:imvtqmd}
\end{eqnarray} 
where we have used that  $\Pi^{00}_{R}(k_0,k)+\Pi^{00}_{A}(k_0,k)$ is real and even in $k_0$. 
The octet propagator appears in \eqref{eq:imvtqmd} 
as $1/(-k_0+i\epsilon)$, which is again a consequence of working at leading order in $E/T\ll 1$
and neglecting rescattering effects. It is because the even part of $1/(-k_0+i\epsilon)$ is proportional to $\delta(k_0)$ that 
only longitudinal gluons contribute, for the chromoelectric dipole interaction due to transverse gluons is proportional to $k_0$. 
Equation~\eqref{eq:imvtqmd} would also follow from expanding eq.~(\ref{eq:imvtqtsimr}) in $r$.  
As in that case we have regulated the infrared divergence in dimensional regularization. 
The contribution from the scale $m_D$ is the one computed in eq.~\eqref{mdcontribrggmd}, 
whose quark and gluon contributions may be disentangled by means of eqs.~\eqref{defmd}, \eqref{eq:rel4q} and \eqref{eq:rel4g}. 
The reason is again that $E/m_D \ll 1$ and we are working at leading order in $E/m_D$. 
The fact that we are working at leading order in $E/T$ and $E/m_D$ is ultimately 
also the reason why $1/m$ effects provide only subleading thermal corrections to the static result.
Summing the contributions coming from the scale $T$ and the scale $m_D$ we get a finite expression 
for the light-quark contribution to the imaginary part of the potential, which, cast in \eqref{defwidth}, provides 
a thermal decay width of the form \eqref{defsigma} with 
\begin{equation}
\Sigma_q(r,q)=\frac{g^4C_Fn_f r^2}{3\pi}\left[\ln\left(\frac{2q}{m_D}\right)-1\right].
\end{equation}
From this it follows that the cross section of a weakly-coupled $1S$ quarkonium state with light quarks from the medium reads
\begin{equation}
\sigma^{1S}_q(q)=\sigma_{cq}\left[\ln\left(\frac{4q^2}{m^2_D}\right)-2\right],
\label{crossquarkmdgge}
\end{equation}
where $\sigma_{cq}$ has been defined in eq.~(\ref{eq:defscq}).

\begin{figure}[ht]
\begin{center}
\includegraphics[width=12cm]{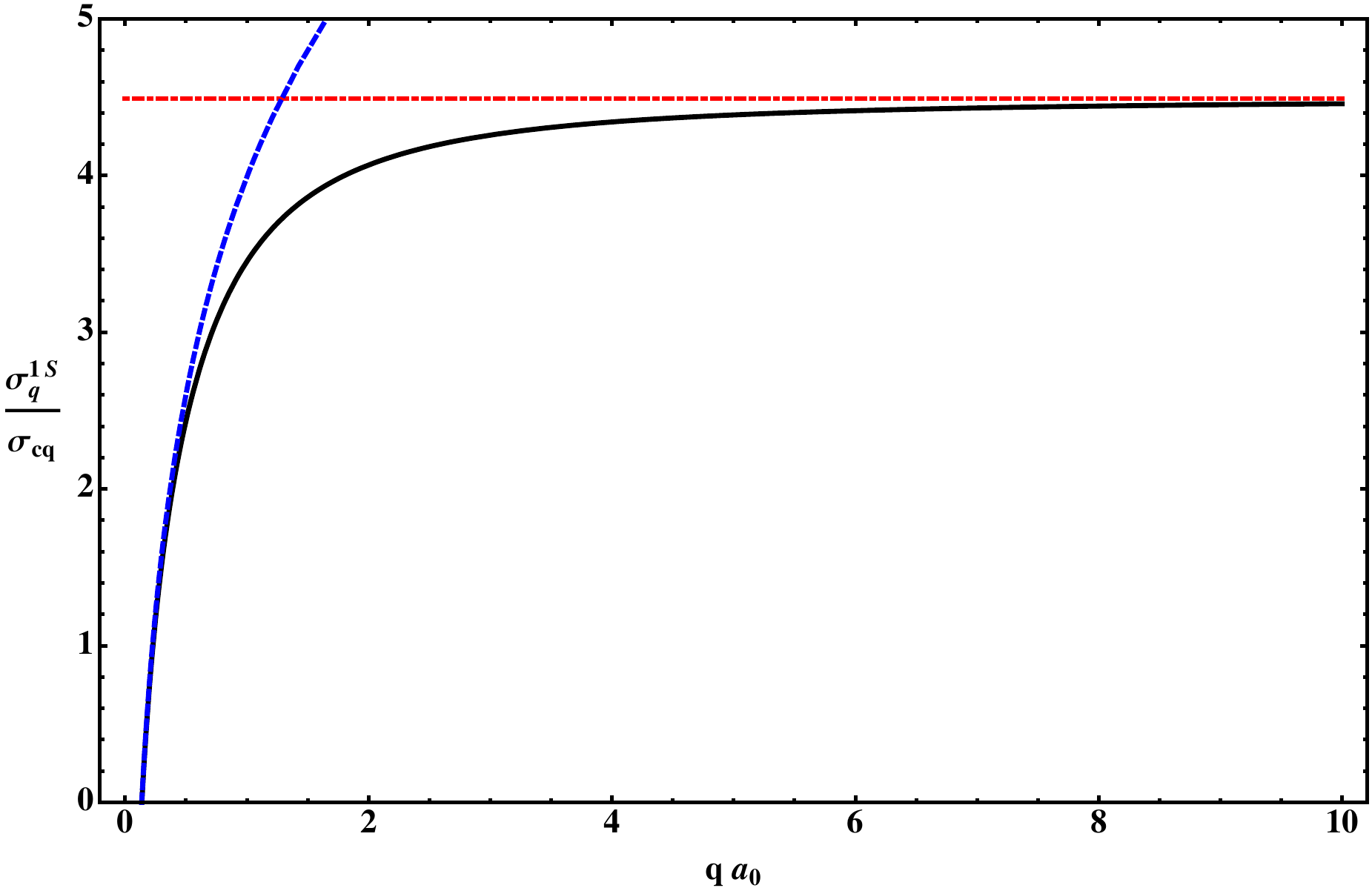}
\end{center}
\caption{Rescaled  dissociation cross sections due to light
  quarks, $\sigma^{1S}_q/\sigma_{cq}$, as a function of the rescaled
  momentum $qa_0$. The dashed blue curve displays the cross section  
  for $mv\gg T\gg m_D\gg E$ given in eq.~\eqref{crossquarkmdgge}, 
  the continuous black curve displays the cross section for $T\sim mv\gg m_D$ 
  given in eq.~\eqref{crossquarktsimr}, and the dot-dashed red curve 
  displays the cross section for $T\gg mv \gg m_D$ given in 
  eq.~\eqref{crossrggmd}. For all the curves we have assumed $m_Da_0=0.1$.}
\label{fig:plotsumq}
\end{figure}

Performing a similar calculation for the part involving the gluon loop, at the scale $T$ we obtain
\begin{eqnarray}
\nn&&\mathrm{Im}\,V^T_g(r)= - \frac{1}{2} \,\mu^{4-D}
\int\frac{d^Dk}{(2\pi)^D}\,\mathrm{Im}\,\left[\frac{1}{-k_0+i\epsilon}k^2 
\frac{r^2}{D-1}g^2\cf 
\frac{i \Pi^{00}_{S,\,g}(k_0,k)}{k^4} \right]\\
\nn&&\hspace{5mm}=-\frac{\pi g^4C_FN_cr^2}{D-1}
\int\frac{\,d^3q}{(2\pi)^3}n_\mathrm{B}(q)(1+n_\mathrm{B}(q))
\mu^{4-D}\int_{2q \ge k}\frac{\,d^{D-1}k}{(2\pi)^{D-1}}\frac{1}{k^3}\left(1-\frac{k^2}{2q^2}+\frac{k^4}{8q^4}\right).\\
&&\label{eq:imvtgmd}
\end{eqnarray}
The contribution from the scale $m_D$ is the gluonic part of eq.~\eqref{mdcontribrggmd};
summing it to eq.~\eqref{eq:imvtgmd} we get the gluonic contribution to the imaginary part of the potential, which, 
cast in \eqref{defwidth}, provides a thermal decay width of the form \eqref{defsigma} with 
\begin{equation}
\Sigma_g(r,q)=\frac{g^4C_FN_c r^2}{3\pi}\left[\ln\left(\frac{2q}{m_D}\right)-1\right].
\end{equation}
From this it follows that the cross section of a weakly-coupled $1S$ quarkonium state with gluons from the medium reads
\begin{equation}
\sigma^{1S}_g(q)=\sigma_{cg}\left[\ln\left(\frac{4q^2}{m^2_D}\right)-2\right],
\label{crossgluonmdgge}
\end{equation}
where $\sigma_{cg}$ has been defined  in eq.~(\ref{eq:defscg}). 

\begin{figure}[ht]
\begin{center}
\includegraphics[width=12cm]{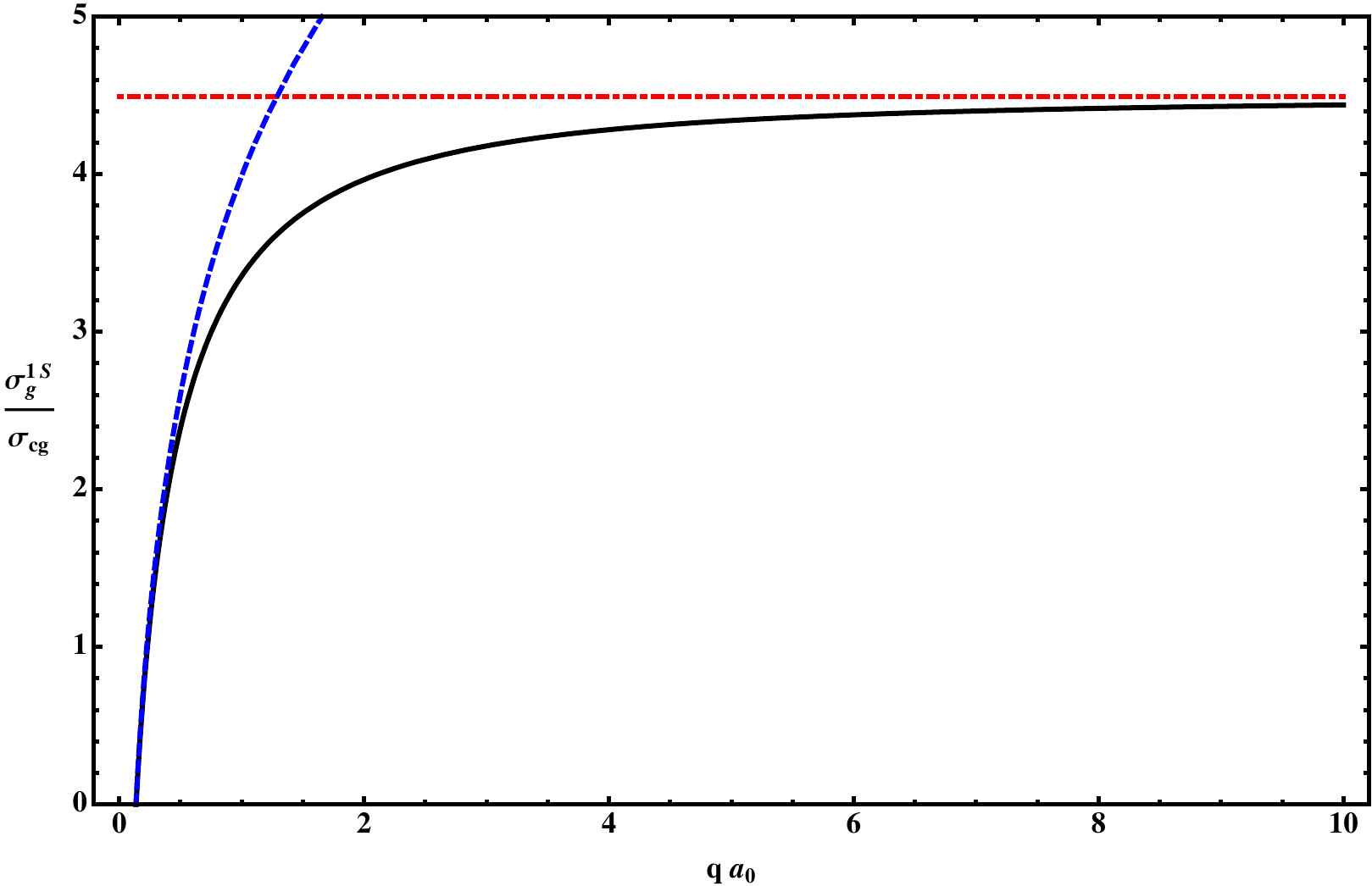}
\end{center}
\caption{Rescaled dissociation cross sections due to gluons, 
  $\sigma^{1S}_g/\sigma_{cg}$, as a function of the rescaled
  momentum $qa_0$. The dashed blue curve displays the cross section  
  for $mv\gg T\gg m_D\gg E$ given in eq.~\eqref{crossgluonmdgge}, 
  the continuous black curve displays the cross section for $T\sim mv\gg m_D$ 
  given in eq.~\eqref{crossgluontsimr}, and the dot-dashed red curve 
  displays the cross section for $T\gg mv \gg m_D$ given in 
  eq.~\eqref{crossrggmd}. For all the curves we have assumed $m_Da_0=0.1$.}
\label{fig:plotsumg}
\end{figure}

We can now compare the results obtained in section~\ref{sec_tsimr} for $T\sim mv\gg m_D$ 
to the two limiting cases: $T \gg mv \gg m_D$, discussed 
in section~\ref{submvggmD}, and $mv \gg T \gg m_D \gg E$, discussed here. 
Physically this means going from temperatures close to the dissociation temperature, $T_d$, 
to temperatures in which all the interactions with the medium can be described 
at leading order by a chromoelectric dipole vertex. 
In~figure~\ref{fig:plotsumq}, we plot the three cross sections with light quarks for $m_Da_0=0.1$. 
The continuous black curve is the result for $T\sim mv \gg m_D$ as given by
eq.~\eqref{crossquarktsimr}: for $qa_0\ll 1$ it is indeed approximated very well 
by the dashed blue line, which is the result for $mv\gg T\gg m_D\gg E$ just obtained 
in eq.~\eqref{crossquarkmdgge}; for $qa_0 \simg 4$ the black line is well approximated 
by the constant, dot-dashed red line, which is the result for $T \gg mv \gg m_D$, 
as given in eq.~\eqref{crossrggmd}.
In figure~\ref{fig:plotsumg}, similar curves show the three cross sections with gluons.

\subsection{Comparison with the literature}
The results of this section bear a direct relation to the \emph{momentum diffusion
coefficient} $\kappa$ of a single heavy quark, first introduced and computed in~\cite{Moore:2004tg}.
According to the field-theoretical definition of~\cite{CasalderreySolana:2006rq}, the diffusion coefficient 
can be written as the time integral of the thermal expectation value of two chromoelectric fields 
linked by Wilson lines stretching along the temporal axis. 
Our eqs.~\eqref{eq:imvtqmd} and \eqref{eq:imvtgmd} correspond indeed to the integral of the 
correlator of two chromoelectric fields up to a factor $r^2$. Equation~(B 13) of~\cite{Moore:2004tg} 
matches the structure of eqs.~\eqref{eq:imvtqmd} and~\eqref{eq:imvtgmd}, while eq.~(B 14) of~\cite{Moore:2004tg} 
contains the same $q$-dependent factor, $[\ln(4q^2/m_D^2)-2]$, that we find in eqs.~\eqref{crossquarkmdgge} and \eqref{crossgluonmdgge}.
Only in this section, when the thermal scales are set between the bound state scales $mv$ and $mv^2$, 
which clearly do not appear in the single quark case, is this direct comparison possible. 
It would then be also possible to use the NLO computation of $\kappa$ in~\cite{CaronHuot:2007gq,CaronHuot:2008uh} 
to obtain some of the order $g$ corrections to the width, while 
bound-state dependent contributions of possibly the same order, like those coming from the expansion
of the octet potential, would need a new dedicated computation.\footnote{
We remark that the NLO calculation of~\cite{CaronHuot:2007gq,CaronHuot:2008uh}, when
applied to heavy quarkonium, shows how the distinction between dissociation by inelastic parton scattering
and gluo-dissociation fails beyond leading order. For instance, the intricate structure of cuts discussed 
in~\cite{CaronHuot:2007gq,CaronHuot:2008uh} includes processes with two light partons in the initial and final state.}

\section{The $mv\gg T\gg E\gg m_D$ case}
\label{sec_eggmd} 
The temperature region $mv\gg T\gg E\gg m_D$ was studied in detail in~\cite{Brambilla:2010vq}. 
This temperature region is technically more difficult to treat 
than the other ones discussed in the paper. One of the reasons is that we find diagrams 
whose imaginary parts contribute both to gluo-dissociation 
and dissociation by inelastic parton scattering. Another reason is that the calculation of these diagrams 
involves not only longitudinal gluons but also transverse gluons.
Since the temperature is smaller than the typical inverse radius of the quarkonium, 
all quarkonium interactions  with the medium are described 
at leading order by chromoelectric dipole vertices.
Hence, also the effects discussed in this section are specific of having an interacting 
$Q\overline{Q}$ system and would be absent in the quasi-free approximation.

Our starting point is the pNRQCD Lagrangian defined in the previous section after 
integrating out the temperature $T$. The reason we can start from there 
is that the Lagrangian of an EFT is only sensitive to the hierarchy of energy scales 
above its ultraviolet cutoff. Therefore integrating out the scales $mv$ and $T$ is 
unaffected by the relation between $m_D$ and $E$. Because the Lagrangian at the scale $T$ 
is the same, the leading contributions to the imaginary part of the static potential 
from the scale $T$ can be read off directly from eqs.~(\ref{eq:imvtqmd}) and (\ref{eq:imvtgmd}). 

The next step consists in integrating over momenta of the order of the binding energy~$E$. 
At this scale, the octet propagator in the diagram of figure~\ref{fig:selfcut2} 
can no longer be taken as $1/(-k_0+i\epsilon)$, for the momentum $k_\mu$ flowing 
in the loop is of the same order as the octet energy. This means that the rescattering 
between the unbound heavy-quarks and their relative motion has to be taken into account. As a consequence,  
the interaction between heavy quarks and transverse gluons, which is proportional 
to $k_0$, does not vanish and contributes to the computation. 
This should be contrasted with what occurs for other energy scales. 

The retarded/advanced propagator of a transverse gluon in Coulomb gauge after HTL resummation reads
\begin{equation}
D^{ij}_{R,A}(k_0,k) = i \frac{\delta^{ij}-\hat{k}^i\hat{k}^j}{(k_0\pm i\epsilon)^2-k^2-\Pi^T_{R,A}\left({k_0}/{k}\right)}\,,
\label{Dtranscoul}
\end{equation}
where $\Pi^T_{R,A}(k_0/k) = (\delta^{ij}-\hat{k}^i\hat{k}^j)\Pi^{ij}(k_0\pm i\epsilon,k)/2$ is 
the HTL retarded/advanced transverse gluon self energy.\footnote{
The explicit expression of  $\Pi^T_{R,A}$ is~\cite{Kalashnikov:1979cy,Weldon:1982aq}
$$
\Pi^T_{R,A}(k_0/k) =  \frac{m_D^2}{2}
\left[\frac{k_0^2}{k^2}-\left(\frac{k_0^2}{k^2}-1\right)\frac{k_0}{2k}
\ln\left(\frac{k_0+k\pm i\epsilon}{k_0-k\pm i\epsilon}\right)\right].
$$
}
It has the following properties. First, $\Pi^T_{R,A}(0)=0$. 
Second, $\mathrm{Im}\,\Pi^T_{R,A}(k_0/k)$ is different from $0$ only for $|k_0/k|<1$.
The fact that the imaginary part of $\Pi^T_{R,A}$ does not vanish only for space-like momenta 
implies that it contributes only to quarkonium  dissociation through inelastic parton scattering.
Finally, we have that $\mathrm{Re}\,\Pi^T_{R,A}\left({k_0}/{k}\simg 1\right) \sim m_D^2$.
The equation $k_0^2-k^2-\mathrm{Re}\,\Pi^T_{R,A}\left({k_0}/{k}\right)=0$ has then a solution 
only for time-like momenta. This solution is called plasmon pole and obeys a special 
dispersion relation with a momentum-dependent mass. For time-like momenta the imaginary part 
of $iD^{ij}_{R,A}$ comes only from the $i\epsilon$ prescription on the plasmon pole. 
Hence the plasmon pole contributes only to the quarkonium gluo-dissociation.
There are two distinct momentum regions where the gluon propagator is nearly singular or singular.
One is the region where $k_0\sim k \sim E$ and $k_0^2-k^2 \sim E^2$.
In this region, which has been called off-shell region in~\cite{Brambilla:2010vq}, 
we can expand the gluon propagator in the transverse gluon self energy.
The other is the region where $k_0\sim k \sim E$ and $k_0^2-k^2 \sim m_D^2$.
In this region, which has been called collinear region in~\cite{Brambilla:2010vq}, 
HTL effects have to be resummed.

The situation at the scale $E$ is therefore the following.
Gluons interact with the $Q\overline{Q}$ pair at leading order through chromoelectric dipole interactions. 
The relevant Feynman diagram is shown in figure~\ref{fig:selfcut2}.
Because at this scale the rescattering of the $Q\overline{Q}$ pair cannot be neglected,  
both longitudinal and transverse gluons contribute to the quarkonium thermal decay. 
Quarkonium may decay either by emitting time-like or light-like gluons, which corresponds 
to cutting the diagram in figure~\ref{fig:selfcut2} along the gluon propagator and 
picking up its pole contribution, or by scattering with partons in the medium.
This last situation corresponds to cutting the diagram in figure~\ref{fig:selfcut2}
along the gluon self-energy diagram and picking up its discontinuity,  which is 
encoded in the symmetric polarization tensor. The momentum of the gluon is in this 
case space-like. According to our definitions the first decay process contributes 
to  quarkonium gluo-dissociation whereas the second one to  dissociation by inelastic parton scattering. 
Both decay processes are intertwined at the scale $E$ and may be disentangled 
only by looking at the time-like or space-like nature 
of the gluon interacting with the $Q\overline{Q}$ pair.
If the gluon interacting with the $Q\overline{Q}$ pair is longitudinal,
then the residue of its plasmon pole contribution is exponentially suppressed 
for momenta $k_0\sim k \sim E \gg m_D$~\cite{Pisarski:1989cs,Blaizot:2001nr}, 
whereas a contribution to the thermal width comes from the imaginary part of 
the longitudinal polarization tensor. This is different from zero only for space-like 
momenta and hence contributes only to quarkonium  dissociation  by inelastic parton scattering.
If the gluon interacting with the $Q\overline{Q}$ pair is transverse, 
then it may contribute either through its plasmon pole or through the imaginary part 
of the transverse gluon self energy. The former case, which may happen only in the 
collinear momentum region for time-like gluon momenta, contributes to quarkonium 
gluo-dissociation. The latter case, which may happen both in the collinear and in the 
off-shell momentum regions for space-like momenta, contributes to  quarkonium  dissociation by inelastic parton scattering.
We will disentangle the gluo- and parton-scattering dissociation contributions 
to the cross section in the following two sections.

\subsection{Gluo-dissociation}
Quarkonium gluo-dissociation was studied in~\cite{Brambilla:2011sg} at leading order in an $m_D/E$ expansion, 
which corresponds to evaluating the diagram in figure~\ref{fig:selfcut2} with a free gluon.
Here we add HTL effects, which amounts at computing the gluo-dissociation width and cross section 
at NLO. It is when HTL effects are taken into account that also parton-scattering dissociation happens.

For the reasons discussed in the previous paragraphs, gluo-dissociation is due to the diagram shown in figure~\ref{fig:selfcut2}
when the gluon is transverse and its momentum time-like.
The relevant contributions have been calculated in appendix~A of~\cite{Brambilla:2010vq}. 
Using those results the gluo-dissociation thermal width at next-to-leading order in $m_D/E$ and $E/T$ reads
\begin{eqnarray}
\Gamma_{nl} &=& \frac{4}{3}\als C_FT \langle n,l\vert r_i(E-h^{(0)}_o)^2r_i \vert n,l\rangle 
+  \frac{2}{3}\als C_F \langle n,l\vert r_i(E-h^{(0)}_o)^3r_i \vert n,l\rangle 
\nonumber\\
&& 
-\frac{\als C_F T m_D^2}{3} \langle n,l\vert r_i\left[\ln\left(\frac{8(E-h^{(0)}_o)^2}{m_D^2}\right)-2\right]r_i \vert n,l\rangle
\,,
\label{Gammagluo}
\end{eqnarray}
where  $h^{(0)}_o = {\bp^2}/{m}+ {\als}/{(2\nc\,r)}$ is the leading-order octet Hamiltonian.
The first term in the right-hand side of~\eqref{Gammagluo} is the leading (zeroth-)order width 
in the $m_D/E$ and $E/T$ expansions: it reproduces the result of~\cite{Brambilla:2011sg} for $T\gg E$. 
The other two terms are the next-to-leading-order corrections.

\begin{figure}[ht]
\begin{center}
\includegraphics[scale=0.9]{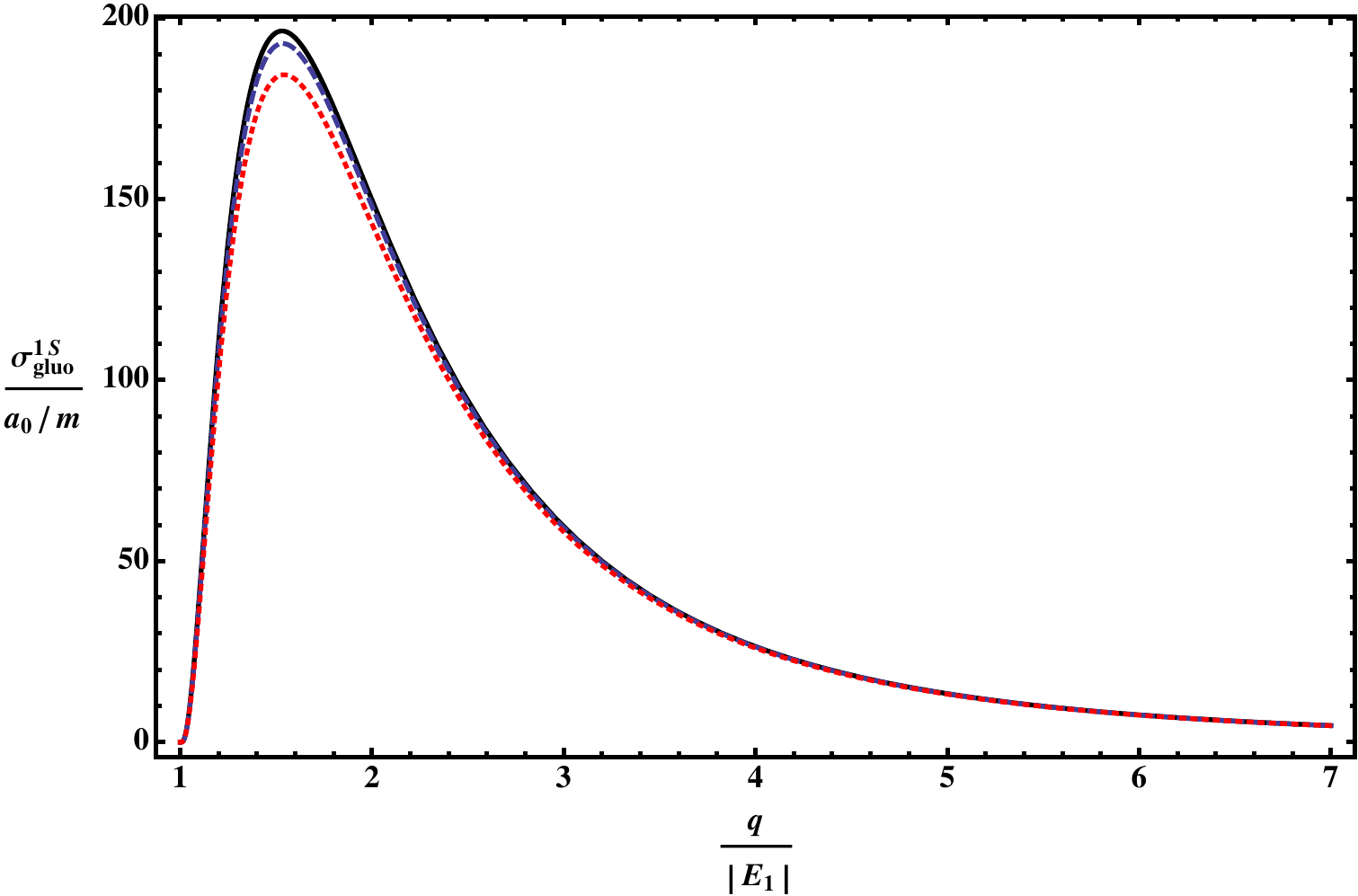}
\end{center}
\caption{$1S$ gluo-dissociation cross section in units of $a_0/m$. The
  continuous black line is the leading-order result, $\sigma_{\rm gluo}^{1S\, (0)}$.
  The dashed blue line shows $\sigma_{\rm gluo}^{1S}$, according to eq.~\eqref{gluonlo},
  for $m_D\,ma_0^2=0.2$ and the dotted red line for $m_D\,ma_0^2=0.5$.}
\label{fig:plotgluo}
\end{figure}

The gluo-dissociation width may be expressed as the convolution 
of a gluon-heavy-quarkonium dissociation cross section in the medium, $\sigma_{\rm gluo}^{nl}$, 
and a gluon distribution function:
\begin{equation}
\Gamma_{nl}= \int_{q_\mathrm{min}}\frac{d^3q}{(2\pi)^3}\,n_{\rm B}(q)\,\sigma_{\rm gluo}^{nl}(q)\,.
\label{eq:fac:gluo}
\end{equation} 
Note that, in contrast to the parton-scattering dissociation case, described by eq.~\eqref{eq:sec2}, there is just one 
parton of the medium involved in the gluo-dissociation process and therefore just one distribution 
function appearing in~\eqref{eq:fac:gluo}. 
Comparing \eqref{eq:fac:gluo} with \eqref{Gammagluo} and expanding the
Bose--Einstein distribution for $T \gg E$, we obtain the gluo-dissociation cross section 
\begin{equation}
\sigma_{\rm gluo}^{nl}(q) = Z({q}/{m_D})\,\sigma_{\rm gluo}^{nl\, (0)}(q)\,,
\label{gluonlo}
\end{equation}
where $\sigma_{\rm gluo}^{nl\, (0)}$ is the leading-order cross section that corresponds to the 
first line in the right-hand side of~\eqref{Gammagluo}. Its explicit expression for a $1S$ Coulombic 
bound state can be found in~\cite{Brambilla:2011sg,Brezinski:2011ju} and reads
\begin{equation}
\sigma_{\rm gluo}^{1S\, (0)}(q) = \frac{\als\cf}{3} 2^{10} \pi^2  \rho  (\rho +2)^2 \frac{E_1^{4}}{mq^5}
\left(t(q)^2+\rho ^2	\right)\frac{e^{\frac{4 \rho}{t(q)}  \arctan
\left(t(q)\right)}}{ e^{\frac{2 \pi  \rho}{t(q)} }-1}\,,
\label{crossEFT}
\end{equation}
where $\rho\equiv 1/(\nc^2-1)$, $t(q)\equiv\sqrt{q/\vert E_1\vert-1}$ and $ E_1 = - m\cf^2\als^2/4$ 
is the energy of the first Bohr level. The absolute value of $E_1$ provides the low-momentum cut-off 
in the integral~\eqref{eq:fac:gluo}. The factor $Z$ can be understood as a wave-function 
normalization of the gluon due to the HTL resummation, it reads
\begin{equation}
Z(x)=1-\frac{1}{4x^2}\left[\ln(8x^2)-2\right]\,.
\end{equation}

The effect of the normalization factor $Z$ on the $1S$ gluo-dissociation cross section 
is shown in figure~\ref{fig:plotgluo}, where the cross section is expressed in units of  
$a_0/m=2/(m^2\cf\als)$ and the gluon momentum in units of $\vert E_1\vert$. 
The plot shows how the HTL resummation results in a global lowering of the cross section.\footnote{
$Z(q/m_D)$ becomes larger than one for $q<2^{-3/2} e\,m_D\approx 0.96\,m_D$. 
However the cross section has a threshold at $q=|E_1|$, and $|E_1|$ is larger than $m_D$ in the assumed hierarchy.}

\subsection{Dissociation by inelastic parton scattering} 
Contributions from the scale $E$  to quarkonium  dissociation  by inelastic parton scattering 
come from 
the different sources that we have analyzed in the introduction of section~\ref{sec_eggmd}. 
We have the contribution from longitudinal gluons and that from transverse gluons.
The contribution from transverse gluons is divided into contributions 
from the collinear region and from the off-shell region, which are separated by a cut-off. 
Only the sum of all these contributions is gauge invariant and cut-off independent.
These different contributions have all been computed:
the contribution from the longitudinal gluons can be found in eq.~(5.17) of~\cite{Brambilla:2010vq};
the contributions from the transverse gluons, of which we have to keep only the contributions coming 
from space-like momenta, can be found in appendix~A of~\cite{Brambilla:2010vq}.

The final light-quark loop contribution to the decay width from the scale $E$ is 
\begin{eqnarray}
\nn\Gamma^{E}_{nl,\,q}&=&-\frac{g^4 C_Fn_f}{3\pi}\langle n,l|r^2|n,l\rangle
\int\frac{\,d^3q}{(2\pi)^3}n_\mathrm{F}(q)(1-n_\mathrm{F}(q))\left[\frac{1}{D-4}-\frac{1}{2}\ln(2\pi)\right.
\\
&&\hspace{7.2cm}\left.+\frac{\gamma_E}{2}-\frac{5}{6}+\ln\left(\frac{m_D}{\mu}\right)\right].
\label{quarkEggmd}
\end{eqnarray}
Adding to it the contribution from the scale $T$, as given by eqs.~\eqref{eq:imvtqmd} and \eqref{defwidth}, 
the divergence cancels and we can cast the decay width in the form \eqref{defsigma} with 
\begin{equation}
\Sigma_q(r,q)=\frac{16 \pi C_Fn_f\,\als^2 \,r^2}{3}\left[\ln\left(\frac{2q}{m_D}\right)+\frac{\ln 2}{2}-1\right].
\label{sigmaqnlo}
\end{equation}
For a $1S$ Coulombic state, the corresponding quark-heavy-quarkonium dissociation cross section then reads 
\begin{equation}
\sigma^{1S}_q(q)= \sigma_{cq} \left[\ln\left(\frac{4q^2}{m_D^2}\right)+ \ln 2-2\right],
\label{sigmaq1Snlo}
\end{equation}
where $\sigma_{cq}$ has been defined in~\eqref{eq:defscq}.

The gluon loop contribution to the decay width from the scale $E$ is the same up to a different colour structure
and different distribution functions: 
\begin{eqnarray}
\nn\Gamma^{E}_{nl,\,g}&=&-\frac{g^4 C_FN_c}{3\pi}\langle n,l|r^2|n,l\rangle
\int\frac{\,d^3q}{(2\pi)^3}n_\mathrm{B}(q)(1+n_\mathrm{B}(q))\left[\frac{1}{D-4}-\frac{1}{2}\ln(2\pi)\right.
\\ 
&&\hspace{7.2cm} \left.+\frac{\gamma_E}{2}-\frac{5}{6}+\ln\left(\frac{m_D}{\mu}\right)\right].
\label{gluonEggmd}
\end{eqnarray}
Adding to it the contribution from the scale $T$, as given by eqs.~\eqref{eq:imvtgmd} and \eqref{defwidth}, 
the divergence cancels and we can cast the decay width in the form \eqref{defsigma} with 
\begin{equation}
\Sigma_g(r,q)=\frac{16\pi C_FN_c\,\als^2\,r^2}{3}\left[\ln\left(\frac{2q}{m_D}\right)+\frac{\ln 2}{2}-1\right].
\label{sigmagnlo}
\end{equation}
For a $1S$ Coulombic state, the corresponding gluon-heavy-quarkonium dissociation cross section then reads 
\begin{equation}
\sigma^{1S}_g(q)= \sigma_{cg} \left[\ln\left(\frac{4q^2}{m_D^2}\right)+ \ln 2-2\right],
\label{sigmag1Snlo}
\end{equation}
where $\sigma_{cg}$ has been defined in~\eqref{eq:defscg}.

The parton-scattering decay width  is of order $\als T \times (m_D/mv)^2$, therefore suppressed by a factor $(m_D/E)^2$ with respect 
to the gluo-dissociation width, which at leading order scales like $ \als T \times (E/mv)^2$. 
The parton-scattering decay width is comparable in size to the next-to-leading-order correction 
to the gluo-dissociation width that appears in the second line of \eqref{Gammagluo}, while the 
next-to-leading-order correction appearing in the first line of \eqref{Gammagluo} is of order 
$\als T \times (E/mv)^2 \times (E/T)$. 
Hence, in the  temperature region $mv\gg T\gg E\gg m_D$,  dissociation  by inelastic parton
 scattering is a subleading 
effect with respect to gluo-dissociation and may be neglected in first approximation.
At the same time we observe that in all other temperature regions examined in the paper we had $m_D \gg E$.
Therefore, in those regions, just the opposite holds and 
 dissociation by inelastic parton scattering is the parametrically 
dominant quarkonium dissociation process.
These observations agree with the early findings of~\cite{Brambilla:2008cx}. The dominance of
dissociation by inelastic parton scattering over gluo-dissociation 
at high temperatures was noticed in~\cite{Grandchamp:2001pf,Grandchamp:2002wp}.

\subsection{Comparison with the literature}
In~\cite{Park:2007zza} parton-scattering and gluo-dissociation have been 
treated in an unified framework. We will highlight some qualitative features 
of that work that are common also to other approaches but that are different 
from the EFT treatment presented here.
The first difference is that the calculation of~\cite{Park:2007zza} uses 
the formula~\eqref{eq:facw} for both the parton-scattering and the gluo-dissociation widths.
We have seen that this is consistent with QCD only in the latter case.
The cross sections used in~\cite{Park:2007zza} have been derived from a Bethe--Salpeter framework in~\cite{Song:2005yd}.
The calculation includes systematically bound-state effects, but with some limitations: 
it is valid in the large $N_c$ limit, hence it neglects rescattering effects of the unbound colour-octet quark-antiquark pair;
it describes the quarkonium interaction with gluons through chromoelectric dipole vertices, 
hence the description holds for gluons whose energy and momentum are smaller than $mv$.
The cross section does not include systematically thermal effects, for it is calculated 
at zero temperature. Constant thermal masses have been added to regulate infrared divergences.
This amounts at a phenomenological tuning: from a QCD perspective one should recall that momentum and temperature-independent masses are 
neither consistent with HTL resummation nor with weak-coupling perturbative calculations.

\section{Conclusions}
\label{sec_concl} 
Quarkonium dissociation through scattering with light partons 
is one of the processes responsible for the thermal decay width of quarkonium 
in a deconfined medium. It is the dominant process for temperatures 
such that the Debye mass, $m_D$, is larger than the binding energy, $E$. 

We have studied this process in a weak-coupling effective field theory framework, 
where quarkonium dissociation through scattering with partons in the medium may be 
related, for momentum transfer larger than $E$, to the imaginary 
part of the potential and the Landau damping phenomenon.
We have shown that in our setting the quasi-free approximation, which consists 
in  approximating the dissociation cross section of the quarkonium with that 
of two free quarks, is never a valid approximation.
In particular, for momentum transfer smaller than or of the same order as 
the inverse radius of the quarkonium, the dissociation cross section in the 
quasi-free approximation is exactly cancelled by bound-state effects.

The parton-scattering dissociation cross section, valid for temperatures such that $m_D$ is much larger than $E$,  
is of the form~\eqref{crossTmv}, with $\Sigma_q$ given in~\eqref{quarkcontribSigmatsimr} for scattering 
with light quarks from the medium and $\Sigma_g$ given in~\eqref{gluoncontribSigmatsimr} for scattering with gluons. 
In the specific case of a Coulombic $1S$ state, the cross sections with gluons and quarks 
are given respectively by eqs.~\eqref{crossquarktsimr} and~\eqref{crossgluontsimr}  and shown by the black curves 
in figures~\ref{fig:plotsumq} and~\ref{fig:plotsumg}. In the region of temperatures where 
$m_D$ is of the order of the inverse radius of the quarkonium, i.e. in the region where 
screening effects are important, $\Sigma_q$ is given by eq.~\eqref{sigmaq} and 
$\Sigma_g$ by eq.~\eqref{sigmag}. In the case of a Coulombic  $1S$ state, 
the cross sections with quarks and gluons are given respectively by eqs.~\eqref{quarkrsimmd} and~\eqref{gluonrsimmd}. 
The impact of the screening on the cross sections is shown in figures~\ref{fig:plottsimr} and~\ref{fig:plottsimrb}.
The parton-scattering dissociation cross section has been computed also for temperatures such that 
$m_D$ is much smaller than~$E$. The light-quark contribution is given in 
eq.~\eqref{sigmaqnlo} with the corresponding cross section for a $1S$ Coulombic state in~\eqref{sigmaq1Snlo},
and the gluon contribution is given in eq.~\eqref{sigmagnlo} with the corresponding cross section for a $1S$ Coulombic state 
given in~\eqref{sigmag1Snlo}. In this temperature regime, dissociation by inelastic parton scattering is, however, a subleading effect.

From the parton-scattering dissociation cross section we may calculate the corresponding 
dissociation width through eq.~\eqref{eq:sec2}. This expression is justified by general 
arguments based on the optical theorem and cutting rules at finite temperature, and 
by the explicit calculations performed in the paper in the different temperature regimes.
Equation~\eqref{eq:sec2} should replace the widely used formula~\eqref{eq:facw}, which 
is justified only in the gluo-dissociation case.

Gluo-dissociation is the process occuring when a sufficiently energetic gluon of the medium 
is absorbed by the quarkonium and dissociates it into an unbound colour-octet $Q\overline{Q}$ pair.
This is the dominant dissociation process in the temperature region where 
$m_D$ is much smaller than $E$, while it is subleading with respect 
to dissociation  by inelastic parton scattering if $m_D$ is much 
larger than $E$. Gluo-dissociation has been 
called singlet-to-octet thermal breakup in the effective field theory literature on 
the subject. We have calculated the gluo-dissociation cross section 
in eq.~\eqref{gluonlo} and thermal width in eq~\eqref{Gammagluo} at next-to-leading order 
in $m_D/E$ and $E/T$. This is currently the most accurate determination of gluo-dissociation 
in weak coupling, whose impact is shown in figure~\ref{fig:plotgluo}.

\acknowledgments

We acknowledge financial support from the DFG cluster of excellence
\emph{Origin and structure of the universe}
(www.universe-cluster.de). This research is supported by the DFG
grant BR 4058/1-1. The work of J.G. was supported by the Natural
Science and Engineering Research Council of Canada and by an Institute
of Particle Physics Theory Fellowship.

\end{document}